\documentclass[iop,twocolumn]{aastex63}
\usepackage{amsmath}
\usepackage{amssymb, amsmath, apjfonts, natbib, color}
\usepackage{multirow}
\usepackage{array}

\hypersetup{
    colorlinks=true,
    linkcolor=blue,
    filecolor=magenta,      
    urlcolor=cyan,
    pdftitle={Overleaf Example},
    pdfpagemode=FullScreen,
    }

\usepackage{textcomp}
\usepackage{gensymb}
\usepackage{hyperref}

\newcommand\xone{x_1}
\newcommand\xtwo{x_2}
\newcommand\xthree{x_3}

\newcommand\Msun{\; {M}_{\odot}}
\newcommand\Lsun{\; {\rm L}_{\odot}}
\newcommand\kms{\; {\rm km}\;{\rm s}^{-1}}
\newcommand\kmsarcmin{\; {\rm km}\;{\rm s}^{-1}\;{\rm arcmin}^{-1}}
\newcommand\kmspc{\; {\rm km}\;{\rm s}^{-1}\;{\rm pc}^{-1}}
\newcommand\pc{\;{\rm pc}}

\newcommand\kpc{\;{\rm kpc}}

\newcommand\freq{\kms\kpc^{-1}}

\newcommand\Myr{\;{\rm Myr}}
\newcommand\Gyr{\;{\rm Gyr}}

\newcommand\MBH{M_{\rm BH}}

\newcommand\C{{\mathcal C}}

\newcommand\simgt{\lower.5ex\hbox{$\; \buildrel > \over \sim \;$}}
\newcommand\simlt{\lower.5ex\hbox{$\; \buildrel < \over \sim \;$}}
\newcommand{\RNum}[1]{\uppercase\expandafter{\romannumeral #1\relax}}
\newcommand\firstcomponent{\rm [O\:{\RNum{3}}]_1}
\newcommand\secondcomponent{\rm [O\:{\RNum{3}}]_2}
\newcommand\oiii{\rm [O\:{\RNum{3}}]}
\newcommand\HI{\rm H\:{\RNum{1}}}

\graphicspath{{./}{figures/}}

\begin{document}

\title{Large-scale Hydrodynamical Shocks as the Smoking Gun Evidence for a Bar in M31}


\renewcommand{\thefootnote}{\fnsymbol{footnote}}
\renewcommand{\thempfootnote}{\fnsymbol{mpfootnote}}

\author{Zi-Xuan Feng\footnotemark[1]}
\affiliation{Shanghai Astronomical Observatory, Chinese Academy of Sciences, 80 Nandan Road, Shanghai 200030, People's Republic of China}
\affiliation{School of Astronomy and Space Sciences, University of Chinese Academy of Sciences, 19A Yuquan Road, Beijing 100049, People's Republic of China}

\author{Zhi Li\footnotemark[1]}
\affiliation{Department of Astronomy, School of Physics and Astronomy, Shanghai Jiao Tong University, 800 Dongchuan Road, Shanghai 200240, People's Republic of China;  email: jtshen@sjtu.edu.cn}
\affiliation{Tsung-Dao Lee Institute, Shanghai Jiao Tong University, Shanghai 200240, People's Republic of China}

\author{Juntai Shen\footnotemark[2]}
\affiliation{Department of Astronomy, School of Physics and Astronomy, Shanghai Jiao Tong University, 800 Dongchuan Road, Shanghai 200240, People's Republic of China;  email: jtshen@sjtu.edu.cn}
\affiliation{Key Laboratory for Particle Astrophysics and Cosmology (MOE) / Shanghai Key Laboratory for Particle Physics and Cosmology, Shanghai 200240, People's Republic of China}
\affiliation{Shanghai Astronomical Observatory, Chinese Academy of Sciences, 80 Nandan Road, Shanghai 200030, People's Republic of China}

\author{Ortwin Gerhard}
\affiliation{Max-Planck-Institut f{\"u}r Extraterrestrische Physik, Giessenbachstrasse, D-85748 Garching, Germany}

\author{R. P. Saglia}
\affiliation{Max-Planck-Institut f{\"u}r Extraterrestrische Physik, Giessenbachstrasse, D-85748 Garching, Germany}
\affiliation{Universit{\"a}ts-Sternwarte München, Scheinerstr. 1, 81679 Munich, Germany}

\author{Matias. Bla\~{n}a}
\affiliation{Max-Planck-Institut f{\"u}r Extraterrestrische Physik, Giessenbachstrasse, D-85748 Garching, Germany}
\affiliation{Institute of Astrophysics, Pontificia Universidad Católica de Chile, Avenida Vicuña Mackenna 4860, 7820436 Macul, Santiago, Chile}

\footnotetext[1]{Contributed equally to this work.}
\footnotetext[2]{Corresponding author: Juntai Shen}

\begin{abstract}
The formation and evolutionary history of M31 are closely related to its dynamical structures, which remain unclear due to its high inclination. Gas kinematics could provide crucial evidence for the existence of a rotating bar in M31. Using the position-velocity diagram of $\oiii$ and $\HI$, we are able to identify clear sharp velocity jump (shock) features with a typical amplitude over $100\kms$ in the central region of M31 ($4.6\kpc \times 2.3\kpc$, or $20\arcmin \times 10\arcmin$). We also simulate gas morphology and kinematics in barred M31 potentials and find that the bar-induced shocks can produce velocity jumps similar to those in $\oiii$. The identified shock features in both $\oiii$ and $\HI$ are broadly consistent, and they are found mainly on the leading sides of the bar/bulge, following a hallmark pattern expected from the bar-driven gas inflow. Shock features on the far side of the disk are clearer than those on the near side, possibly due to limited data coverage on the near side, as well as obscuration by the warped gas and dust layers. Further hydrodynamical simulations with more sophisticated physics are desired to fully understand the observed gas features and to better constrain the parameters of the bar in M31.

\end{abstract}

\section{Introduction}

Although M31 is the nearest large spiral galaxy to the Milky Way at a distance of $785 \;\pm\; 25\kpc$ \citep{mcc_etal_05}, its exact location on the Hubble tuning fork diagram is still unclear, which is key to understand its formation and evolution. The debates have a quite long history that probably started from \citet{lindbl_56} who first claimed that M31 is a barred galaxy based on the twist of central isophotes. The isophotal twist cannot be reproduced by an axisymmetric stellar distribution \citep{stark_77}. Later studies argued that unbarred galaxies can also have twisted inner isophotes as they can originate from triaxial bulges, not necessarily bars \citep{stark_77, zar_lo_86, ber_etal_88, ger_etal_89, men_etal_10, cos_etal_18}, which left the true morphology of M31 a puzzle. In addition, the strong enhancement of star formation between $2-4$ Gyr ago \citep{wil_etal_15} could be linked to a recent single merger event \citep{ham_etal_18}. The age-velocity dispersion relation in the stellar disk \citep{bha_etal_19} suggested that a merger occurred $\sim3-4$ Gyr ago with an estimated mass ratio of 1:5. The merger would largely enhance the stellar velocity dispersion in the disk and affect the dynamical evolution of M31. After the merger event, another star formation burst in the whole bulge region happened near $\sim$ 1 Gyr ago, which might be caused by the secular evolution \citep{don_etal_16}.

While it is difficult to identify whether there is a bar based on the photometry of a highly inclined M31 disk, numerical simulations have provided more insights into the properties of the inner stellar structures. \citet{ath_bea_06} first used $N$-body models to reproduce the twist of central isophotes and the boxy shape seen in the near-infrared band \citep{beaton_etal_07}. They concluded that M31 has both a classical bulge and a bar whose major axis deviates from that of the disk by about $20^{\circ}$. The scenario is updated by \citet{bla_etal_17} who constructed $N-$body models for M31 to analyse the boxy-shape and the tilted velocity field \citep{opitsc_16}. The authors conclude that the classical bulge and the box/peanut bulge (BPB) in M31 contribute about one-third and two-thirds of the total stellar mass in the central part, respectively. \citet{blana_etal_18} further constructed made-to-measure (m2m) models for M31, using constraints from $3.6\;{\mu}m$ photometry \citep{bar_etal_06} and stellar kinematics \citep{opitsc_etal_18}. The models evolved from $N-$body buckled bar models. The BPB results from a buckled bar in their best model, with a half-length of $\sim4\kpc$ and a pattern speed of $40 \;\pm\; 5\freq$. They also checked a model with a very low pattern speed and found it does not match the data as nicely as the more rapidly rotating models. Moreover, \citet{sag_etal_18} found that the stellar metallicity is enhanced along the proposed bar structure. Based on the m2m models in \citet{blana_etal_18}, \citet{gaj_etal_21} constrained the 3D distribution of metallicity and $\alpha$-enrichment using observed $[Z/H]$ and $[\alpha/Fe]$ maps \citep{sag_etal_18}, and an X-shaped metallicity distribution is found in the bulge region. These results imply that M31 is a barred galaxy.

On the other hand, evidence from gas kinematics in M31 also hints for a bar. Large non-circular motions of gas have been identified in $\HI$ \citep{bri_bur_84, chemin_etal_09} and CO \citep{loinar_95,loinar_99,nieten_etal_06}. In addition, $\oiii$ emission from ionized gas shows a twisted zero-velocity curve in the central bulge region \citep[]{opitsc_etal_18} , and its morphology suggests a tilted spiral pattern with a lower inclination angle compared to the stellar disk \citep{opitsc_etal_18}. All of these are typical features seen in barred galaxies \citep{jacoby_etal_85, emsell_06, kuzio_etal_09, fat_etal_05}, although there are alternative explanations \citep[e.g. a head-on collision between M31 and M32 proposed by][non-circular motions of gas caused by a triaxial bulge]{block_etal_06}. Nevertheless, the overall shape of position-velocity diagrams (PVDs) of $\oiii$ in \citet{opitsc_etal_18} is similar to what has been observed in barred galaxies \citep{bur_ath_99, mer_kui_99}. 

One of the most characteristic gaseous features in typical barred galaxies is a pair of dust-lanes on the leading side of the bar \citep{athana_92}. Dust-lanes are generally associated with shocks, which result in sharp velocity jumps in gas kinematical maps, and thus can be identified by integral-field unit (IFU) data \citep[e.g.][]{opitsc_etal_18}. 

The goal of the current paper is to search and identify such shock (velocity jump) features from the observations. It should be noted that the shocks we are searching for are "large-scale" shocks \citep[e.g. of the type proposed by][]{robert_69,rob_etal_79} which are generally extended over a few kiloparsecs, rather than those "small-scale" shocks due to local turbulence of interstellar medium or supernova feedbacks. If the positions and properties of velocity jumps are similar to those expected from bar-driven shocks, this would provide independent, strong evidence for the existence of a bar in M31.

The paper is organized as follows: \S 2 describes the data of $\oiii$ and $\HI$ in M31. \S 3 discusses the criteria we use to identify shock features on PVDs of $\oiii$ and $\HI$. In \S 4 we present the results, and compare the shock features in $\oiii$ and $\HI$. We also present a map showing the positions of identified shocks. In \S 5 we compare the shock features in $\oiii$ with the bar-driven shocks in hydrodynamical simulations. \S 6 mainly discusses the potential physical reasons for the asymmetry in the shock features between the far side and near side of M31. We briefly summarize our findings in \S 7. 

\begin{figure*}[ht!]
\includegraphics[width=\textwidth]{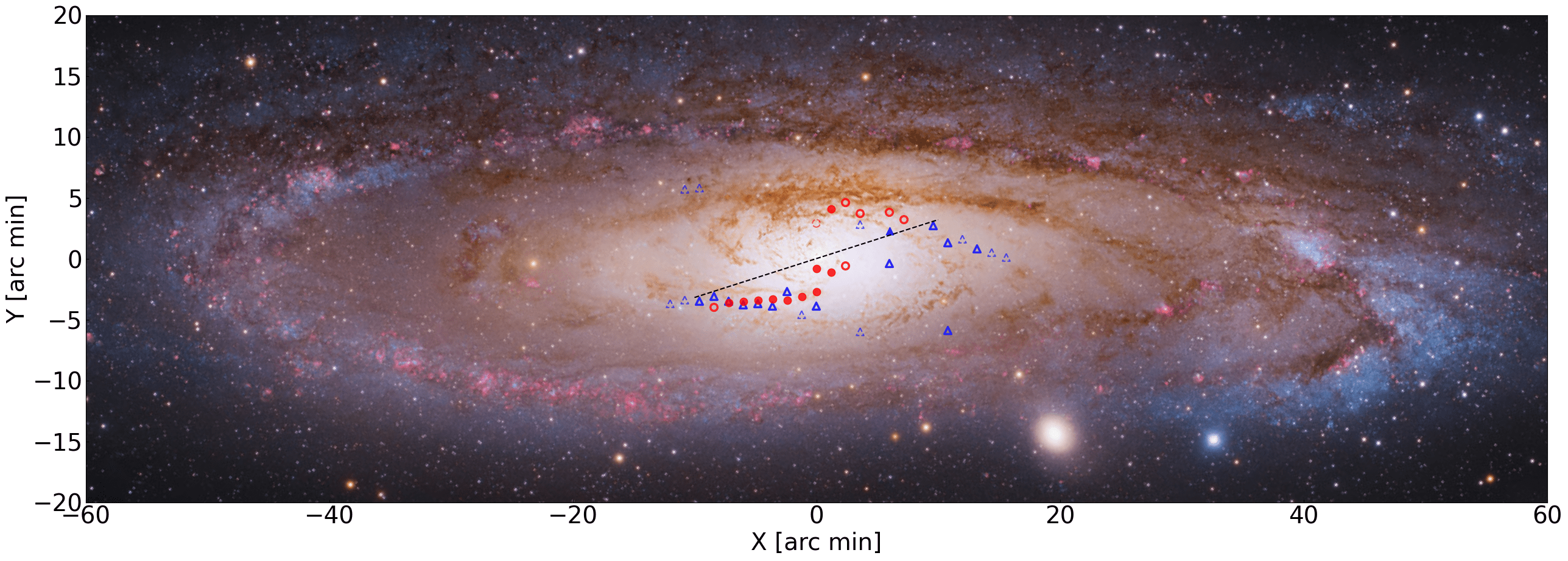}
\caption{Identified shock positions of $\oiii$ (red circles) and $\HI$ (blue triangles) superposed on the optical image of M31. The background image is taken from \url{https://apod.nasa.gov/apod/ap220119.html}. The image is credited to Subaru (NAOJ), Hubble (NASA/ESA), Mayall (NSF) and processed by R. Gendler and R. Croman. X-axis and Y-axis are taken along the major and minor axes of the stellar disk of M31, respectively. Solid, open, and dashed markers represent the Class I, Class II, and Class III shock features, respectively (see Table.~\ref{tab:criteria_shock_feature}). The black dashed line indicates the major axis of the bar in \citet{blana_etal_18}.}
\label{fig:shock_on_optical_image}
\end{figure*}

\section{Observational Data}

The main results of the current paper are summarized in Fig.~\ref{fig:shock_on_optical_image}, which shows the positions of the possible bar-driven shocks superposed on the optical image of M31. The distance to M31 is adopted to be $785 \;\pm\; 25\kpc$ \citep{mcc_etal_05}, so $1\arcmin$ corresponds to $228\pc$. 

We use data from \citet{opitsc_etal_18} and \citet{chemin_etal_09} to study the gas kinematics in M31. \citet{opitsc_etal_18} carried out an IFU survey with the VIRUS-W instrument at the McDonald Observatory, which contains emission lines of $\rm H\beta$, $\oiii$, and $\rm [N\:{\RNum{1}}]$. Their observation covers the inner bulge region of M31 (20$\arcmin \times 10\arcmin$). \citet{chemin_etal_09} observed the 21 cm emission using the Syntheses Telescope at the Dominion Radio Astrophysical Observatory. Their $\HI$ survey covers the whole M31 disk. 

Multiple gas components with different velocities are found over half of the total bins in \citet{opitsc_etal_18}, among which $\oiii$ has the strongest flux. The components of $\oiii$ with higher and lower velocities are labeled as $\firstcomponent$ and $\secondcomponent$, respectively. Multiple gas components are expected when the line of sight passes through different gas streams, which could be caused by a bar \citep{kim_etal_12a} or a collision between M31 and its satellite galaxy M32 \citep{block_etal_06}. We use the main (higher-velocity) component of $\firstcomponent$ to present the PVDs. We discuss the possible origin of the two components in \S\ref{sec:twooxy3}.

Similar to $\oiii$, $\HI$ observations by \citet{chemin_etal_09} show multiple velocity components in their spectra. The $\HI$ disk is extended and starts to warp beyond $\sim 20 \kpc$ \citep{new_eme_77,hender_79}. The $\HI$ component with a lower velocity is likely from the warped $\HI$ layer in the outer disk \citep{bri_bur_84,bri_sha_84}. \citet{chemin_etal_09} shows that gas from the warped region dominates the $\HI$ emission in the central $20\arcmin$, producing shallow linear structures on their PVDs. The warp may obscure the inner $\HI$ disk with similar velocities on PVDs. To find the main $\HI$ component that best represents the disk rotation, \citet{chemin_etal_09} selected the component with the largest velocity relative to the galactic systemic velocity while rejecting isolated faint features. This main $\HI$ component excludes the lower-velocity features that originate in the outer warp as well as the isolated faint features that possibly come from extra-planar gas (e.g. high-velocity clouds). We therefore use this main component of $\HI$ to identify $\HI$ shock features in the disk.

\begin{figure*}[t!]
\includegraphics[width=\textwidth]{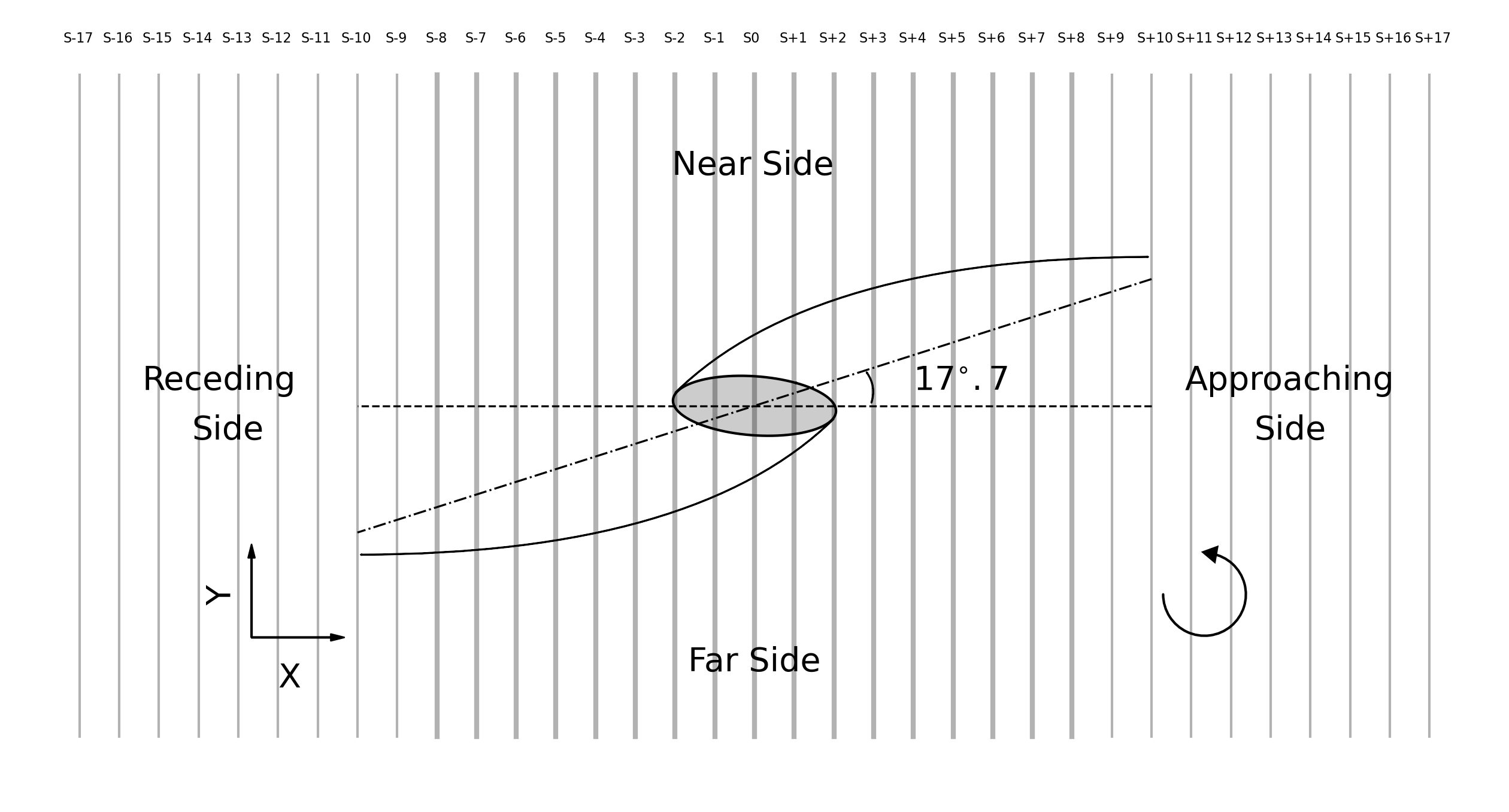}
\caption{Schematic plot of the shocks in a counter-clockwise rotating bar under the viewing angle of M31. The shocks are expected to locate on the leading side of the bar. The major axis of the bar (dash-dotted line) is deviated from the major axis of the disk (dashed line along horizontal direction) by $17.7^{\circ}$. X-axis and Y-axis are taken along the disk major and minor axis, respectively. The vertical solid lines represent the locations of pseudo-slits. The thick and thin vertical lines indicate the slits covering both $\oiii$ and $\HI$ data and those covering only $\HI$ data, respectively. Text on the top labels the pseudo-slits for later reference. S0 is along the disk minor axis. Note that the shaded ellipse represents a projected inner circular gas disk, but not the bar of M31.}

\label{fig:scheme_shock}
\end{figure*}

\section{Construction of Pseudo-slits and Shock Identification}
\label{sec:identification_shocks}

Shocks on the leading side of the bar are commonly found in simulations \citep{rob_etal_79,athana_92} and observations \citep{sandag_61,san_bed_94}. The gas velocity component perpendicular to shocks has an abrupt change, producing sharp velocity jump features on the PVDs. In Fig.~\ref{fig:scheme_shock}, we plot the schematic view of the possible shocks under the viewing angles of M31. The position angle of the projected bar major axis (on the sky frame) from \citet{blana_etal_18} is $55.7^{\circ}$, deviating from $PA_{disk} = 38^{\circ}$ by $17.7^{\circ}$. If M31 is a barred galaxy as suggested by \citet{blana_etal_18}, we expect that shocks appear on the leading side of the bar and extend roughly to the bar ends. Thus, we position pseudo-slits perpendicular to the disk major axis so that the slits cut through the shocks nearly perpendicularly. The standard width of the slit is chosen to be $1.2\arcmin$ which is about twice larger than the spatial resolution of $\HI$ ($43.75\arcsec$). The slits cover the inner $20\arcmin$ where shocks are expected.

The large-scale bar-driven shocks generally produce sharp velocity jumps on PVDs, for which we try to search in this work. For each PVD, the goal is to identify the positions and amplitudes of shock features. \citet{canny_86} detected edges in signals by convolving data with a derivative of Gaussian, and it is now commonly used in 2D image edge detection. We modify their algorithm to detect step-function shaped and $\delta-$function shaped velocity jumps. The steps and instructions of our procedures are described as follows.

\emph{Requirement of sharpness.}
The amplitude of velocity jumps $\Delta V$ as well as the spatial extent $\Delta Y$ determines the sharpness of a shock feature. Simulations by \citet{ath_bur_99} and \citet{kim_etal_12a} have shown that typical bar-driven shocks have $\Delta V > 100-150\kms$ within $\sim 100-200 \pc$ (after projection $200 \times \cos 77^{\circ} = 45 \pc$). We expect the spatial extents of observed velocity jumps to be wider than those in simulations for several reasons: 1. Sharpest velocity jumps are expected when the slits cut shocks perpendicularly, which may not be the case shown in Fig.~\ref{fig:scheme_shock}. 2. Resolution of the observation is usually lower than that of simulations. 3. Dust extinction and local turbulence could blur the velocity field and produce less clear shock features. Considering these effects, we aim to find shock features showing $\Delta V > 100\kms$, $\Delta Y < 0.6\arcmin$ on PVDs, which corresponds to a velocity gradient over $100\kms/0.6\arcmin = 167\kmsarcmin$. We use a larger $\Delta Y$ of 2$\arcmin$ to identify $\HI$ shock features, which is around three times the spatial resolution (43.75$\arcsec$) of $\HI$. Compared to those of $\oiii$, the shock features of $\HI$ have a smaller velocity gradient due to lower spatial resolution.

\begin{figure}[t!]
\includegraphics[width=\columnwidth]{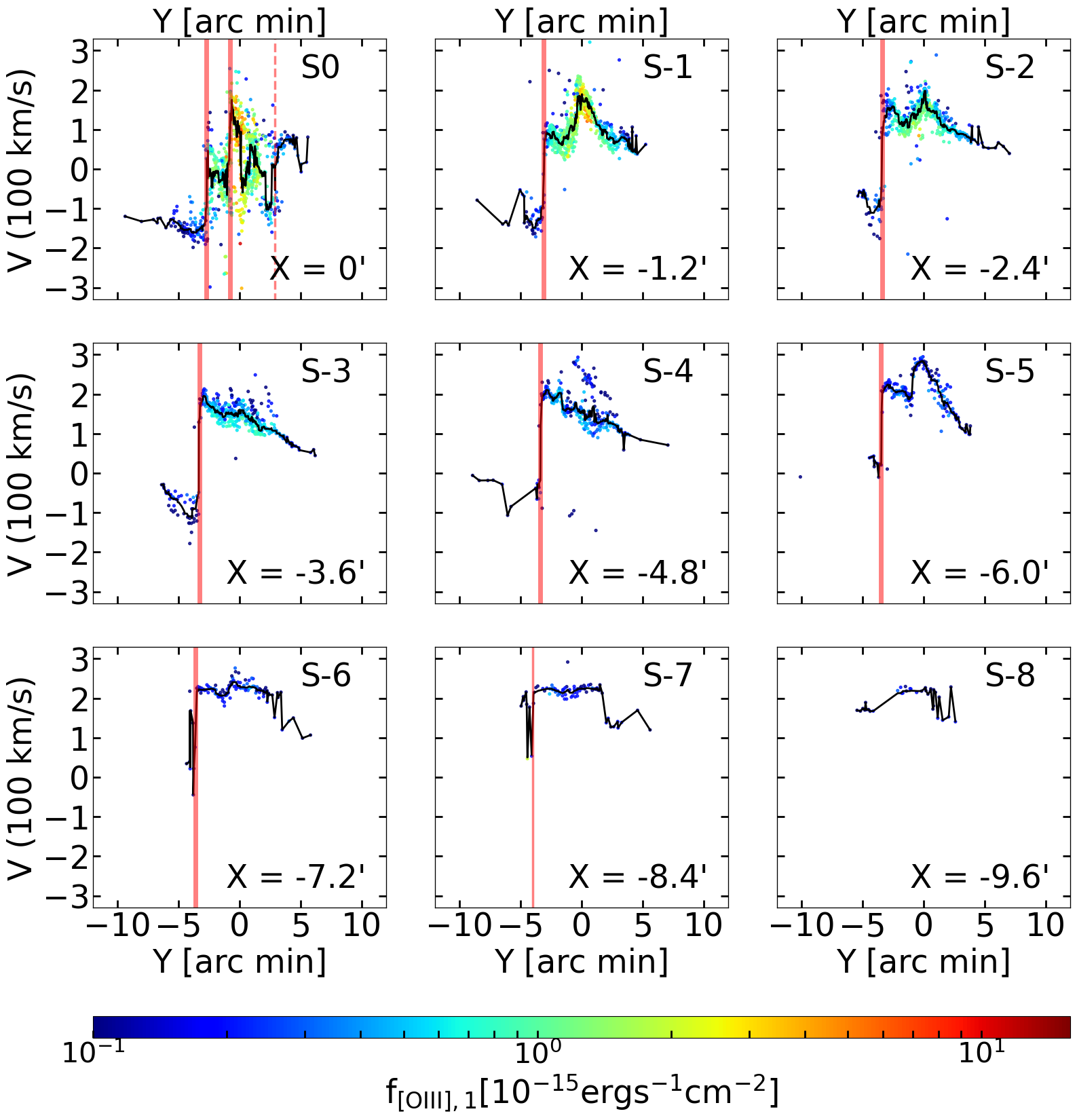}
\caption{PVDs of $\firstcomponent$ color-coded with flux density on the receding side of M31. The data is from \citet{opitsc_etal_18}. Each panel corresponds to one pseudo-slit in Fig.~\ref{fig:scheme_shock}. Coloured points represent the velocities of $\firstcomponent$. Black curves represent the boxcar smoothed result of the coloured points. Thick, thin, and dashed red lines indicate the positions of Class I, Class II, and Class III shock features, respectively. The shock features are clear mostly inside 7.2$\arcmin$ and they are mainly on the far side (negative Y values, below the disk major axis).}
\label{fig:PVDs_component_1_receding_side}
\end{figure}

\begin{figure}[t!]
\includegraphics[width=\columnwidth]{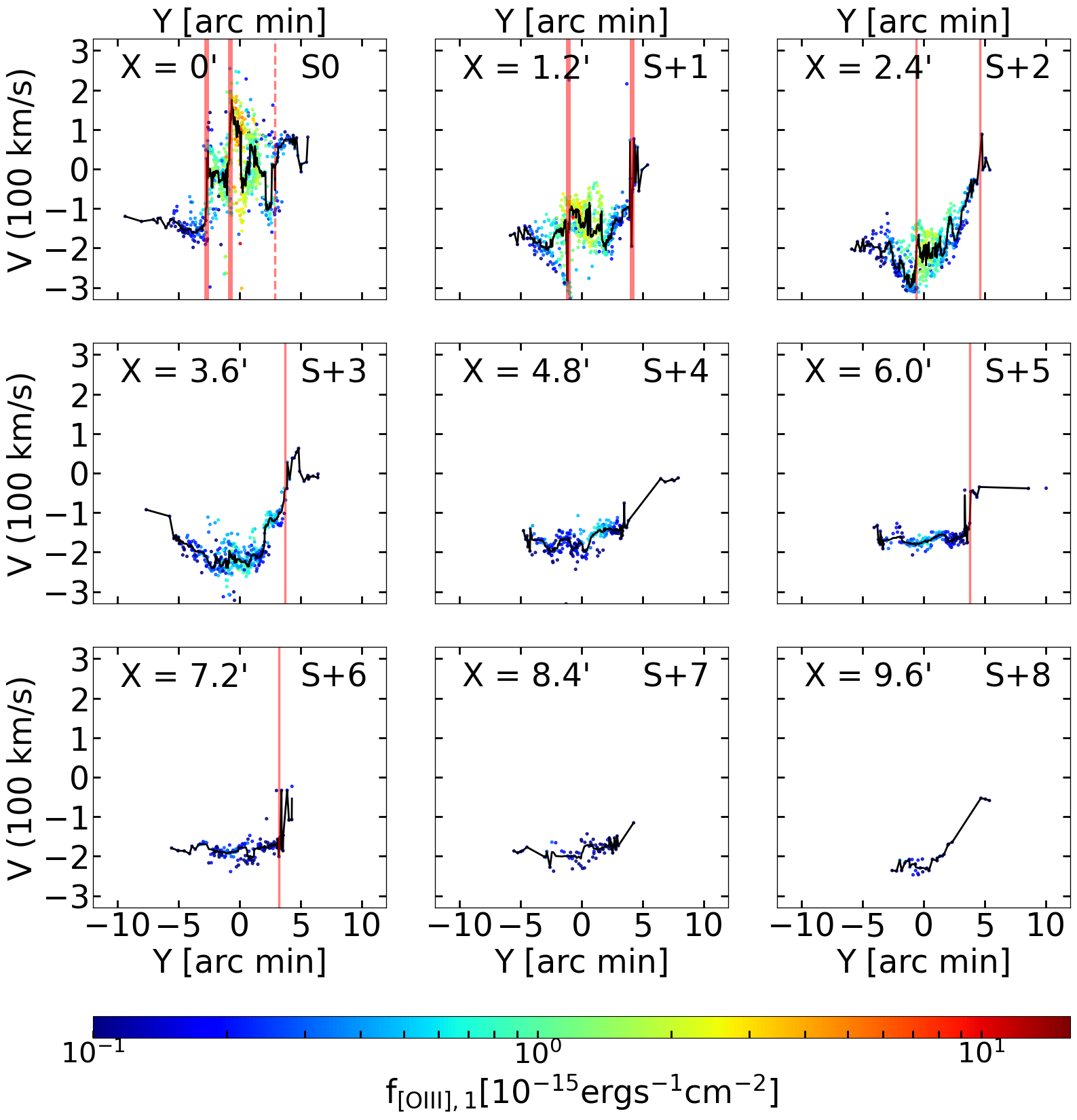}
\caption{Same as in Fig.~\ref{fig:PVDs_component_1_receding_side} but on the approaching side. The shock features are mainly on the near side (positive Y values, above the disk major axis).}
\label{fig:PVDs_component_1_approaching_side}
\end{figure}

\emph{Smoothing}. We use boxcar smoothing to reduce noises on PVDs. Each observed data point is replaced by the median\footnote{We also tested that using median gives sharper shock features than using mean, and the results of the flux density-weighted mean do not differ much from the non-weighted ones.} of its adjacent 15 points for PVDs of $\oiii$. Our tests show that the number of points of 15 is good enough for showing both strong and weak shock features of $\oiii$. We also tested the number as large as 21, the shock features with $\Delta V > 150\kms$ remain robust, but those with small $\Delta V$ are too weak to identify in this case. For shock features close to the boundaries in a few PVDs of $\oiii$ (especially on the near side), the data coverage might be incomplete to show shock features clearly. Therefore we avoid smoothing the 9 points and 15 points close to the far side and near side boundary of $\oiii$ PVDs, respectively. These numbers are empirically chosen to give a more regular shock position pattern. Black lines in Figs.~\ref{fig:PVDs_component_1_receding_side} and \ref{fig:PVDs_component_1_approaching_side} indicate the smoothed $\oiii$ curve. We find that shock features in $\HI$ are more sensitive to the smoothing parameter than that of $\oiii$. We therefore replace each point of $\HI$ main component by the median of its adjacent 5 points to avoid over-smoothing.

\emph{Locating a jump feature.}
Given a curve showing several velocity jumps, we would like to detect jump features automatically. Our procedure first creates a window of 0.6$\arcmin$ (2$\arcmin$ for $\HI$) wide and use it to cover a selected part of the curve. Then it convolves the data within the window using a derivative of Gaussian to highlight the position of the velocity jump, similar to the edge-detection algorithm in \citet{canny_86}. The derivative of the Gaussian operator has the form:

\begin{equation}
y = -\dfrac{x}{s^2}\exp{\dfrac{-x^2}{2\sigma^2}}.
\end{equation}
Here $s = 3$ is a scaling factor and $\sigma = 4$ represents the dispersion of the Gaussian. Vertical lines in Figs.~\ref{fig:PVDs_component_1_receding_side} and \ref{fig:PVDs_component_1_approaching_side} indicate the positions of selected shock features using this method. Tests of the edge-detection algorithm in identifying step-function shaped and $\delta$-function shaped jump features are given in Appendix \ref{appendix:A}.

\emph{Identification.} We calculate the velocity difference within 0.6$\arcmin$ (2$\arcmin$ for $\HI$) for each point on the boxcar smoothed curve. Since we want to find sharp velocity jumps, we focus on regions showing velocity differences over 100$\kms$. Our algorithm targets the region with the largest velocity difference, and returns one shock position using the method above. Then it moves to regions with smaller velocity differences, which departs from the identified shock features by at least 1.5$\arcmin$. We repeat this process several times until all of the velocity jumps over 100$\kms$ are found. The criteria to identify shock features is summarized in Table~\ref{tab:criteria_shock_feature}.

\emph{Classification of shock features.}
We classify the gas velocity jump features into three classes according to the likeliness that a bar-driven shock is present based on $\Delta V$.

$-$ Class I: if the feature shows a velocity jump over $170\kms$, it is classified as a gas feature "very likely" being a shock.

$-$ Class II: if the feature shows a velocity jump between $125\kms$ and $170\kms$, it is classified as a gas feature "likely" being a shock.

$-$ Class III: if the feature shows a velocity jump between $100\kms$ and $125\kms$, it is classified as a feature "possibly" being a shock.

\begin{table}
\caption{Criteria to identify shock (velocity jump) features.}
    \begin{center}
    \begin{tabular}{|m{1.8em}|m{6.7em}|m{5.8em}|m{7.2em}|}
        \hline
        {Class} & {$\Delta V$ within $\Delta Y$ ($\kms$)} & {Possibility of being shocks} & {Cross-comparison} \\
        \hline
        {I} & {$> 170$} & {Very likely} & \multirow{3}{7.5em}{Similar positions on PVDs of $\firstcomponent$ and $\HI$} \\
        \cline{1-3}
        {II} & {$125 \sim 170$} & {Likely} & {} \\
        \cline{1-3}
        {III} & {$100 \sim 125$} & {Possible} & {} \\
        \hline
    \end{tabular}
    \end{center}
\tablenotetext{}{$\Delta Y < 0.6\arcmin$ for $\oiii$ shock features and $\Delta Y < 2\arcmin$ for $\HI$ shock features. Note that the flat part of rotation curve of M31 is around 250 $\kms$.}
\label{tab:criteria_shock_feature}
\end{table}

The $\Delta V$ values in Table 1 are empirically chosen.  Previous observations and simulations of barred galaxies give a rough reference value of shock velocity jump $\Delta V \ga 100\kms$ \citep[e.g.][]{ath_bur_99}, which we choose to be the threshold $\Delta V$ of shock features. We have tested that minor variations of the $\Delta V$ ranges do not change the main result. We expect that positions of Class I and Class II shock features in $\firstcomponent$ are similar to those of $\HI$, except for central regions due to the lack of $\HI$.

\section{Results}
\label{sec:result}

\subsection{Shock features on PVDs in [OIII]}
\label{sec:shock_features_first_component}

Figs.~\ref{fig:PVDs_component_1_receding_side} and \ref{fig:PVDs_component_1_approaching_side} present the PVDs of $\firstcomponent$ and the boxcar smoothed result (black curve) on the receding ($X$ < 0) and approaching ($X$ > 0) side of M31. Negative and positive $Y$ represent the far and near side of M31. Overall, shock features are clearer on the receding side (i.e. Fig.~\ref{fig:PVDs_component_1_receding_side}), showing Class I shock features (thick red lines) in most panels. Class I shock features appear in S0 at $Y \approx -2.7\arcmin$ and $-0.8\arcmin$, together with a Class III shock feature (red dashed line) at $Y \approx 2.9 \arcmin$ (near side). In Fig.~\ref{fig:PVDs_component_1_receding_side}, the clearest shock features are shown in slits (S-1, S-2, S-3, S-4). Positions of Class I shock features shift downwards from $Y \approx -3.1\arcmin$ to $-3.6\arcmin$ as we go from S-1 to S-6. Further out, $\Delta V$ of Class I shock features decreases and the shock features turn into Class II (thin red line) in S-7.

\begin{figure}[t!]
\includegraphics[width=\columnwidth]{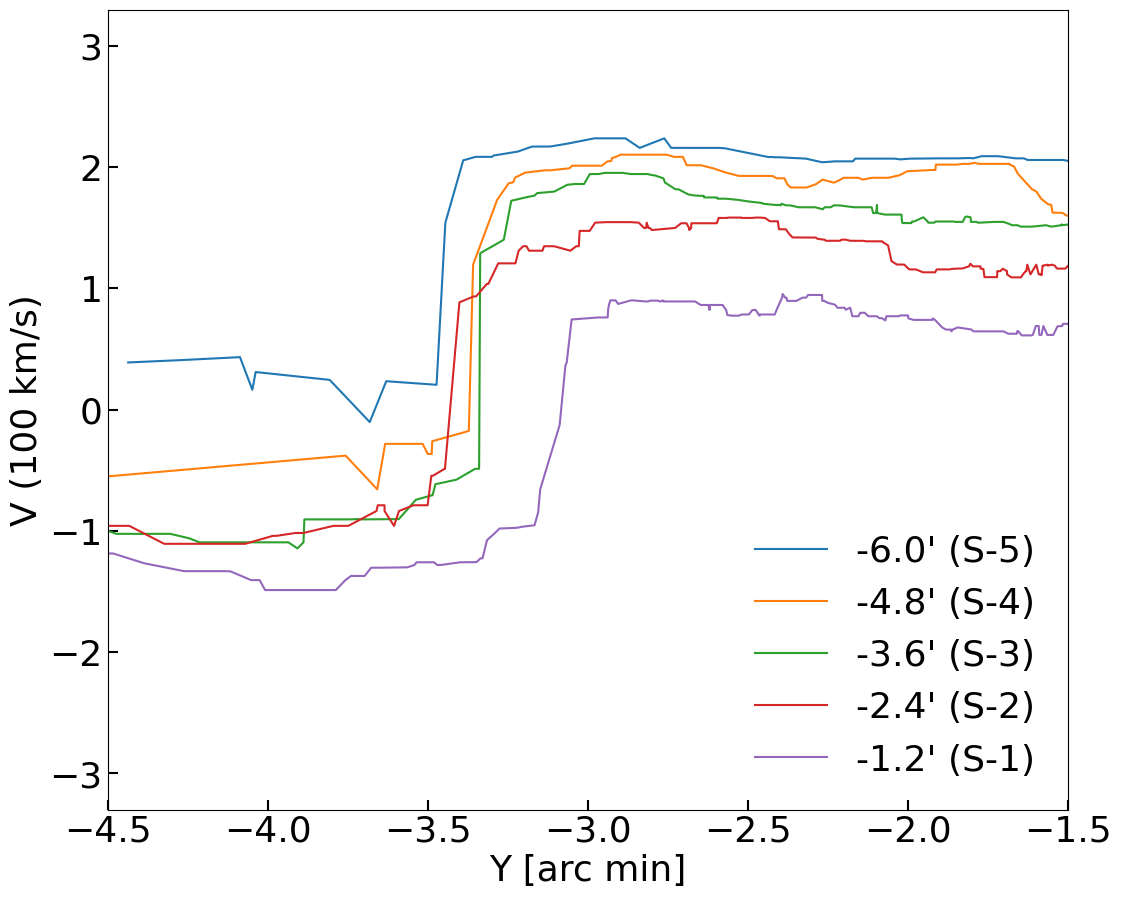}
\caption{Sharp shock features of $\firstcomponent$ on the receding side of M31. Each curve corresponds to a Class I shock feature within slits S-1 to S-5 of Fig.~\ref{fig:PVDs_component_1_receding_side}. The color indicates the $X$ positions of the slits on the disk major axis.}
\label{fig:identified_shocks_component_1_receding_side}
\end{figure}

On the approaching side, Class I and Class II shock features show up in most panels, mainly between $Y \approx 3.2\arcmin$ and $4.6\arcmin$. Using the method described in \S\ref{sec:identification_shocks}, we extract the clearest shock features of $\firstcomponent$ on the receding side, which is shown in Fig.~\ref{fig:identified_shocks_component_1_receding_side}. Color indicates the $X$ positions of the slits. Each curve corresponds to one panel in Fig.~\ref{fig:PVDs_component_1_receding_side}. We do not show the shock features along S-6 and S-7 slits in Fig.~\ref{fig:identified_shocks_component_1_receding_side} for they are too close to the boundary of data coverage. The shock features are found between $Y \approx -3.1\arcmin$ and $-3.5\arcmin$, and shift upwards by $\approx 150 \kms$ as the slit moves from S-1 to S-5.

\subsection{Potential shock features in the HI data}
\label{sec:shock_features_HI}

We also try to cross-validate the shock features in $\firstcomponent$ with archival $\HI$ data. The coverage of $\firstcomponent$ extends to only $\sim500\arcsec$, which is smaller than the projected bar length of $\sim600\arcsec$ in \citet{blana_etal_18}. Therefore we cannot check the regions near the bar ends using only $\firstcomponent$. However, $\HI$ data have several weaknesses: First, the point spread function (PSF) is very large and thus the signal is more smeared. Secondly, a strong warp of $\HI$ disk exists in M31. Line-of-sight velocities of gas in the inner $\HI$ disk are contaminated by the gas in the outer $\HI$ warp. Thirdly, there is an absence of $\HI$ within the central 5$\arcmin$.    

\begin{figure}[t!]
\includegraphics[width=\columnwidth]{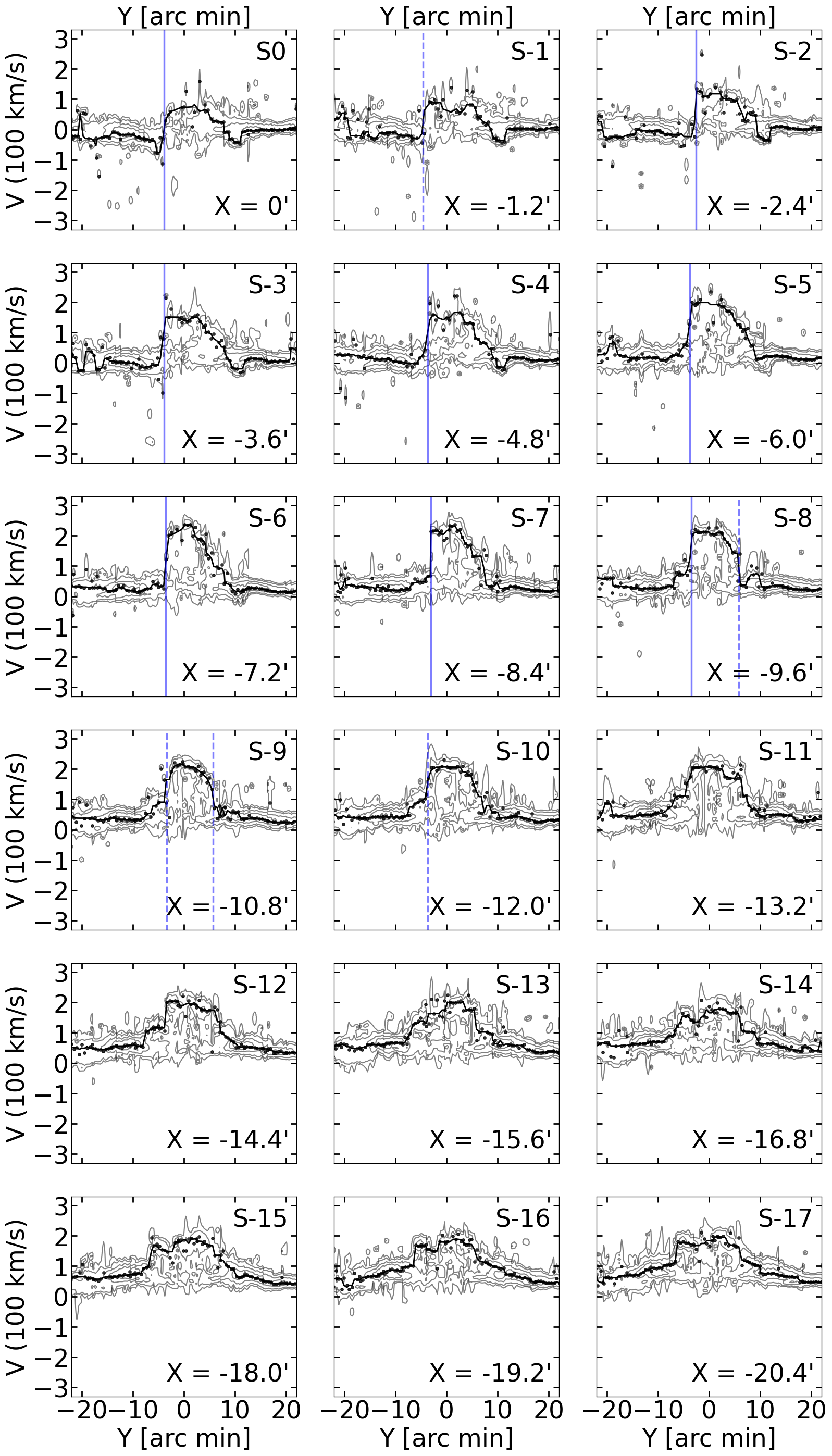}
\caption{PVDs of $\HI$ on the receding side of M31. The data is from \citet{chemin_etal_09}. Each panel corresponds to one pseudo-slit in Fig.~\ref{fig:scheme_shock}. Contours indicate the integrated emission of $\HI$. Black points indicate the velocity of the main component of $\HI$. Black curves represent the boxcar smoothed result of the black points. The thin blue solid line and blue dashed line indicate the positions of Class II and Class III shock features of $\HI$, respectively. Shock features are mainly on the far side (negative Y values, below the disk major axis). Low-velocity features lie horizontally on each panel, which likely come from the warped gas disk.}
\label{fig:PVDs_HI_receding_side}
\end{figure}

\begin{figure}[t!]
\includegraphics[width=\columnwidth]{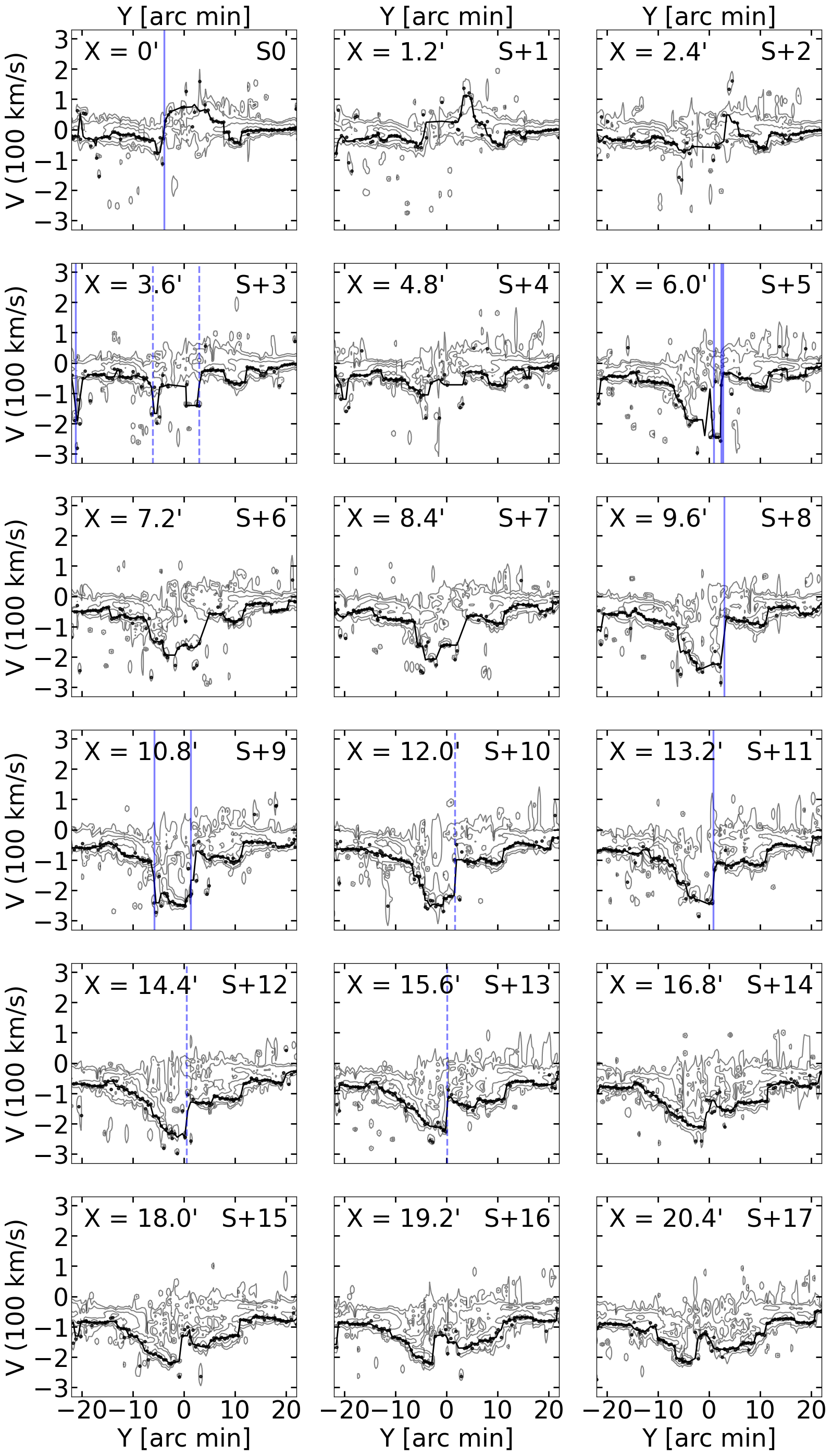}
\caption{Same as in Fig.~\ref{fig:PVDs_HI_receding_side} but on the approaching side. Thick, thin, and dashed blue lines indicate the positions of Class I, Class II, and Class III shock features, respectively.}
\label{fig:PVDs_HI_approaching_side}
\end{figure}

Figs.~\ref{fig:PVDs_HI_receding_side} and \ref{fig:PVDs_HI_approaching_side} show the PVDs of $\HI$ on the receding ($X < 0$) and approaching side ($X > 0$) of M31, respectively. We present the integrated $\HI$ emission with contours and the velocity of the main component of $\HI$ with black points. The black curves represent the boxcar smoothed result of the $\HI$ main component. Overall $\HI$ emission is composed of two parts with different origins and velocities. The gas features with smaller velocities distributing in most regions are possibly caused by the warp in the outer gas disk. The gas features with larger velocities showing beyond the warp features originate from the inner disk. Although emission from the $\HI$ warp obscures parts of shock features with small velocities, Class II shock features (thin blue lines) show up clearly in the main component of $\HI$ in most panels of Fig.~\ref{fig:PVDs_HI_receding_side}. Positions of Class II shock features are found between $Y = -2.7\arcmin$ and $Y = -3.9\arcmin$ on the far side of M31. Further out, $\Delta V$ of Class II shock features decreases, and the shock features turn into Class III (blue dashed lines) in S-9, S-10. On the approaching side, a Class I shock feature (thick blue line) and several weak shock features show up in panels (S+3, S+5, S+8) of Fig.~\ref{fig:PVDs_HI_approaching_side}, but in other regions the gas features with large velocities are quite clumpy and do not show clear shock features. Further out, Class II and III shock features appear near the disk major axis ranging from S+9 to S+13.

We make a cross-comparison of the PVDs of $\HI$ and $\firstcomponent$. The identified shocks distribution of these two tracers are quite similar, hinting for a common origin that is probably due to large-scale bar dynamics. PVDs of S-3, S-4, S-5, S-6 in Fig.~\ref{fig:PVDs_component_1_receding_side} and Fig.~\ref{fig:PVDs_HI_receding_side} illustrate that positions of Class I shock features of $\firstcomponent$ and Class II shock features of $\HI$ are similar. Further out ($X < - 7.2\arcmin$), Class II shock features of $\HI$ have profiles similar to that of $\firstcomponent$ though at slightly different $Y$ positions. Class III shock features are mainly on the near side of M31. There are panels where shock features in $\firstcomponent$ are clearer than in $\HI$, as well as panels where the opposite is true. When we overlay PVDs of $\firstcomponent$ and $\HI$ the shock features are usually easier to recognize.

Fig.~\ref{fig:PVDs_oiii_HI_ShockFeature[-11,11]} compares the PVDs of $\HI$ with several clearest shock features of $\firstcomponent$. The first panel is same as Fig.~\ref{fig:identified_shocks_component_1_receding_side} and it shows the overall pattern of $\firstcomponent$ shock features on the receding side. Other panels show the comparison between $\HI$ and $\firstcomponent$ shock features for each slit. In Fig.~\ref{fig:PVDs_oiii_HI_ShockFeature[-5,-0.5]} we show the same comparison but on a smaller spatial scale. Overall shock features of $\HI$ coincide with $\firstcomponent$, especially at $X = -1.2\arcmin$ and $X = -3.6\arcmin$. The $\HI$ features are quite clumpy, therefore determining the exact shock positions of $\HI$ is not easy. It is possible that the shocks in $\HI$ are not shifted from those in $\firstcomponent$, but obscured or hidden by some missing clumps instead. For example, the shock profile appears to be incomplete at $X = -2.4\arcmin$. If there were a clump near $Y \sim -3.5\arcmin$ with a velocity around $-100\kms$, the shock feature would have been clearer to see.

\subsection{The map of shock positions}
\label{sec:map_shock_position}

We plot the positions of shock features of $\firstcomponent$ (red circles) and $\HI$ (blue triangles) in Fig.~\ref{fig:position_shock}. We use the solid markers to represent Class I shock features, open markers for Class II shock features, and dashed markers for Class III shock features. The solid line in Fig.~\ref{fig:position_shock} represents the bar in \citet{blana_etal_18} with a projected bar angle of $17.7^{\circ}$ and a projected half-length of $2.3 \kpc$. We also show the fitted $3.6\;\mu m$ isophote (dashed) in \citet{blana_etal_18} that is closest to the bar ends. 

The shock features are found mainly on the leading side of the bar and this is consistent with our expectation of bar-driven shocks. In general, shock features are clearer on the far side of M31 than on the near side. The shock positions of $\HI$ and $\oiii$ are very similar for $X$ between -7.2$\arcmin$ and -3.6$\arcmin$. Further out, Class II shock features of $\HI$ extend to the bar ends and turn into Class III shock features with smaller velocity jumps. In the central region of M31, $\oiii$ shows shock features that do not show in $\HI$. Such difference could be due to the lack of $\HI$ in the central regions.

\begin{figure}[t!]
\includegraphics[width=\columnwidth]{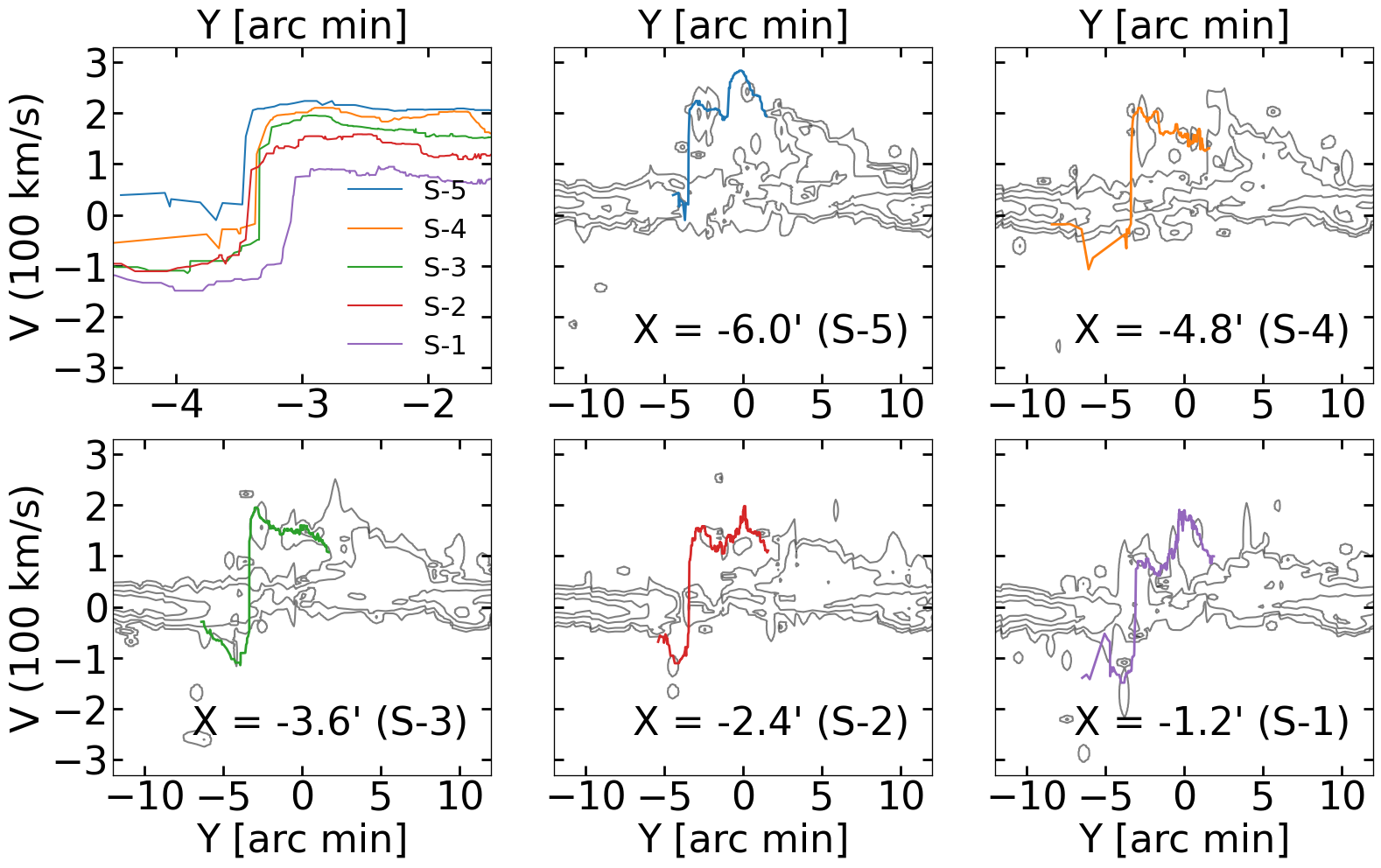}
\caption{\textit{First panel}: same as Fig.~\ref{fig:identified_shocks_component_1_receding_side}. \textit{Other panels}: comparison of shock features of $\firstcomponent$ (curves) and PVDs of $\HI$ (contours) on the receding side. The slit numbers are also labeled.}
\label{fig:PVDs_oiii_HI_ShockFeature[-11,11]}
\end{figure}

\begin{figure}[t!]
\includegraphics[width=\columnwidth]{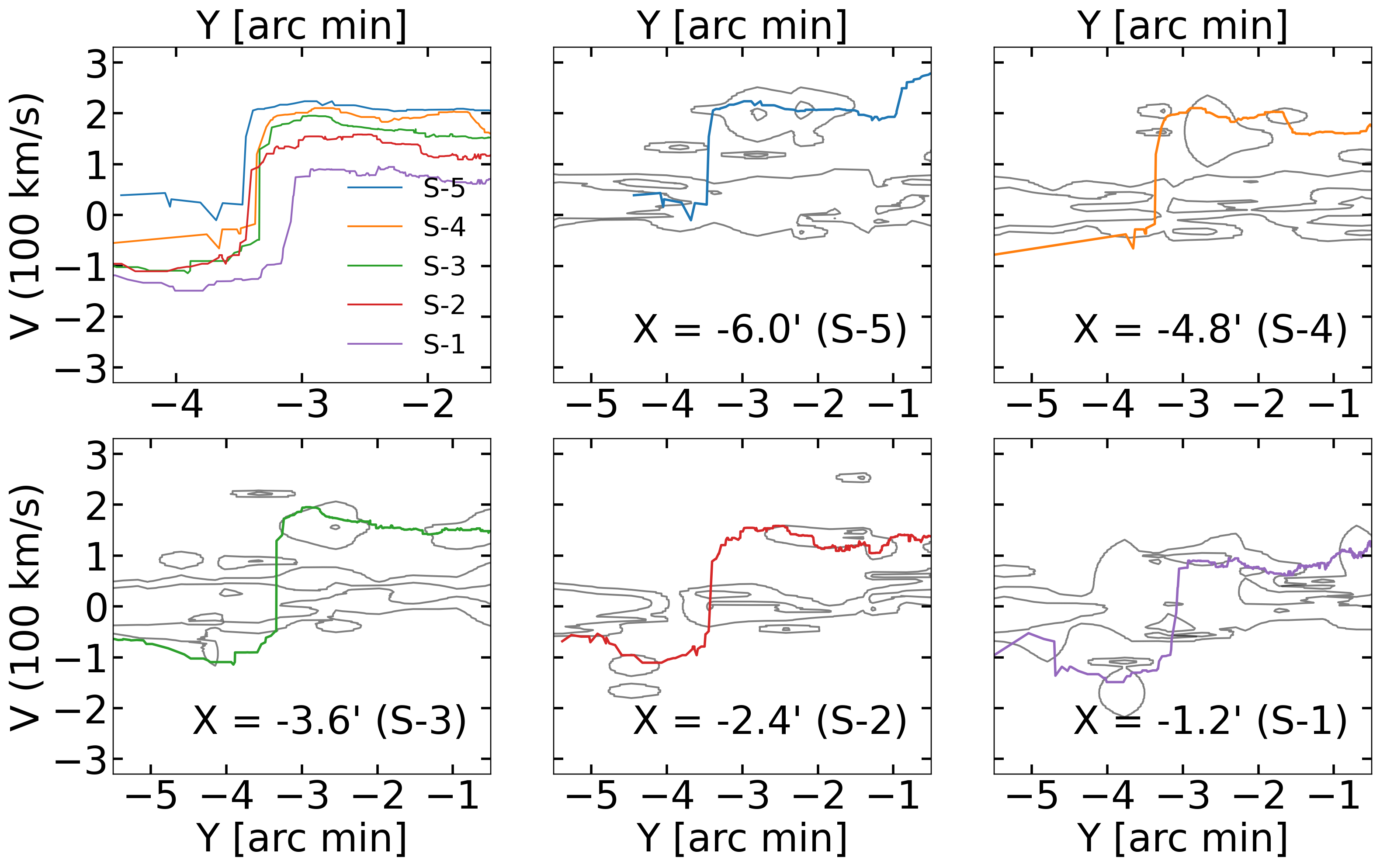}
\caption{\textit{First panel}: same as Fig.~\ref{fig:identified_shocks_component_1_receding_side}. \textit{Other panels}: same as Fig.~\ref{fig:PVDs_oiii_HI_ShockFeature[-11,11]} but on a smaller spatial scale.}
\label{fig:PVDs_oiii_HI_ShockFeature[-5,-0.5]}
\end{figure}

Shock features are, however, absent near $X\sim5\arcmin$. The observation of $\firstcomponent$ mainly covers the bulge region with $|Y| < 5\arcmin$, but it covers only $|Y| < 4\arcmin$ at $X = 5\arcmin$. Class I and Class II shock features in $\firstcomponent$ are found at larger $Y$ distances on the near side than the far side by $\sim 0.5\arcmin$. It is possible that shock features do exist, but extend beyond the data coverage near $X\sim5\arcmin$. 

\section{Simulated gas flow versus the observed shocks}
\subsection{Hydrodynamical simulation}
We also make simple isothermal 2D gas models in a constrained M31 potential to compare with the observed shock features. The simulations here are for illustrative purposes only, but not meant to match all the details of shocks. 

We solve Euler equations in the initial frame using the grid-based MHD code {\tt Athena++} \citep{sto_etal_20}. We adopt a uniform Cartesian grid with a resolution of 4096 $\times$ 4096 covering the simulation domain with length $L = 15 \kpc$ in each direction. The setting corresponds to a grid spacing of ${\Delta}x = {\Delta}y = 7.3\pc$.

\begin{figure}[t!]
\includegraphics[width=\columnwidth]{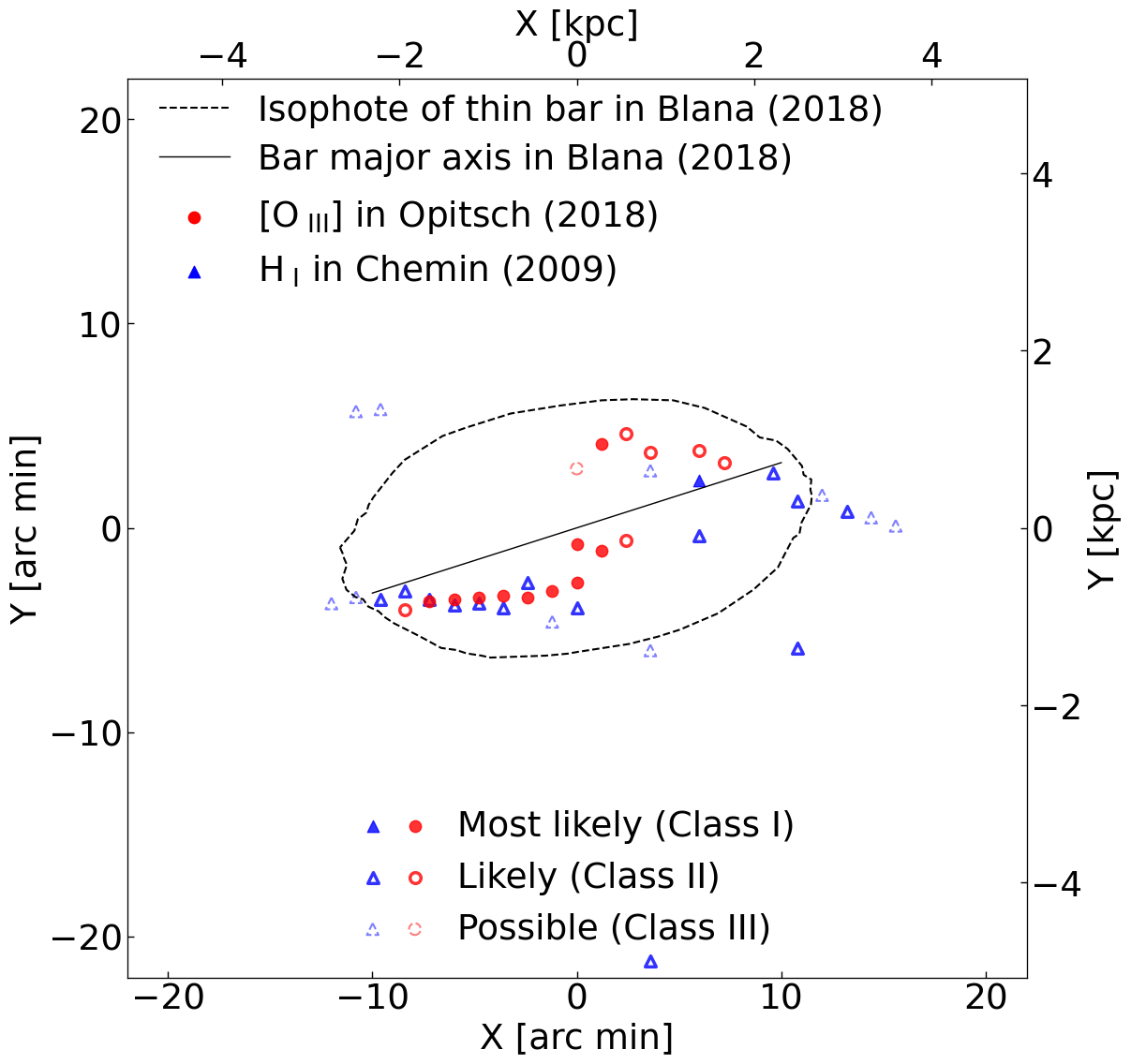} 
\caption{Spatial map of $\firstcomponent$ and $\HI$ shock features. The solid line represents the bar major axis in \citet{blana_etal_18} with a projected half-length of $2.3\kpc$. The dashed line represents the fitted $3.6\;\mu m$ band isophote in \citet{blana_etal_18} that is closest to the end of the bar. Red points and blue triangles indicate the shock positions in $\firstcomponent$ and in $\HI$ data in \S\ref{sec:result}, respectively. Solid, open and dashed markers represent Class I, Class II and Class III shock features.}
\label{fig:position_shock}
\end{figure}

For simplicity, we use the isothermal equation of state $P = \Sigma \;c_s^2$, here $P$, $\Sigma$ denote the gas pressure and gas surface density, respectively. The effective sound speed $c_s$ describes the turbulent properties of the gas. \citet{kim_etal_12a} systematically explores the effects of $c_s$ on gas substructures in barred potentials. They found that models with larger $c_s$ are more perturbed, and produce off-axis shocks closer to the bar major axis and smaller nuclear rings. Previous gas dynamics study suggests that the isothermal assumption can explain many observed gas features. The gas dynamics simulation of \citet{li_etal_16} explained various observed features on the Galactic $l-v$ diagram. \citet{wei_etal_01b,wei_etal_01a} ran hydrodynamical simulations of gas flow in the barred galaxy NGC 4123. By matching the non-circular motions near the dust lane, \citet{wei_etal_01b,wei_etal_01a} suggested a high-mass stellar disk and a fast-rotating bar in NGC 4123. 

We set the initial surface density distribution of the gas disk to an exponential profile $\Sigma_{gas}(R) = \Sigma_{0}\;{\rm exp}(-R/R_{d})$, here $\Sigma_{0} = 76 \Msun\pc^{-2}$ and $R_{d} = 6.0\kpc$, which results in a total mass of 1.2$\times 10^{10}\Msun$. The initial azimuthal rotation velocity of the gas disk is set to balance the azimuthally averaged gravitational force. We input linear growth of the non-axisymmetric force and gradually ramp up the barred potential in one bar rotation period. We accomplish this by increasing the fraction of bar potential from 0 to 1.0 and decreasing the fraction of axisymmetrized bar potential from 1.0 to 0 linearly with time in $T_{grow} = 2\pi/{\Omega_{b}}$, similar to previous studies \citep[e.g.]{athana_92,kim_etal_12a}.

\begin{figure}[ht!]
\includegraphics[width=\columnwidth]{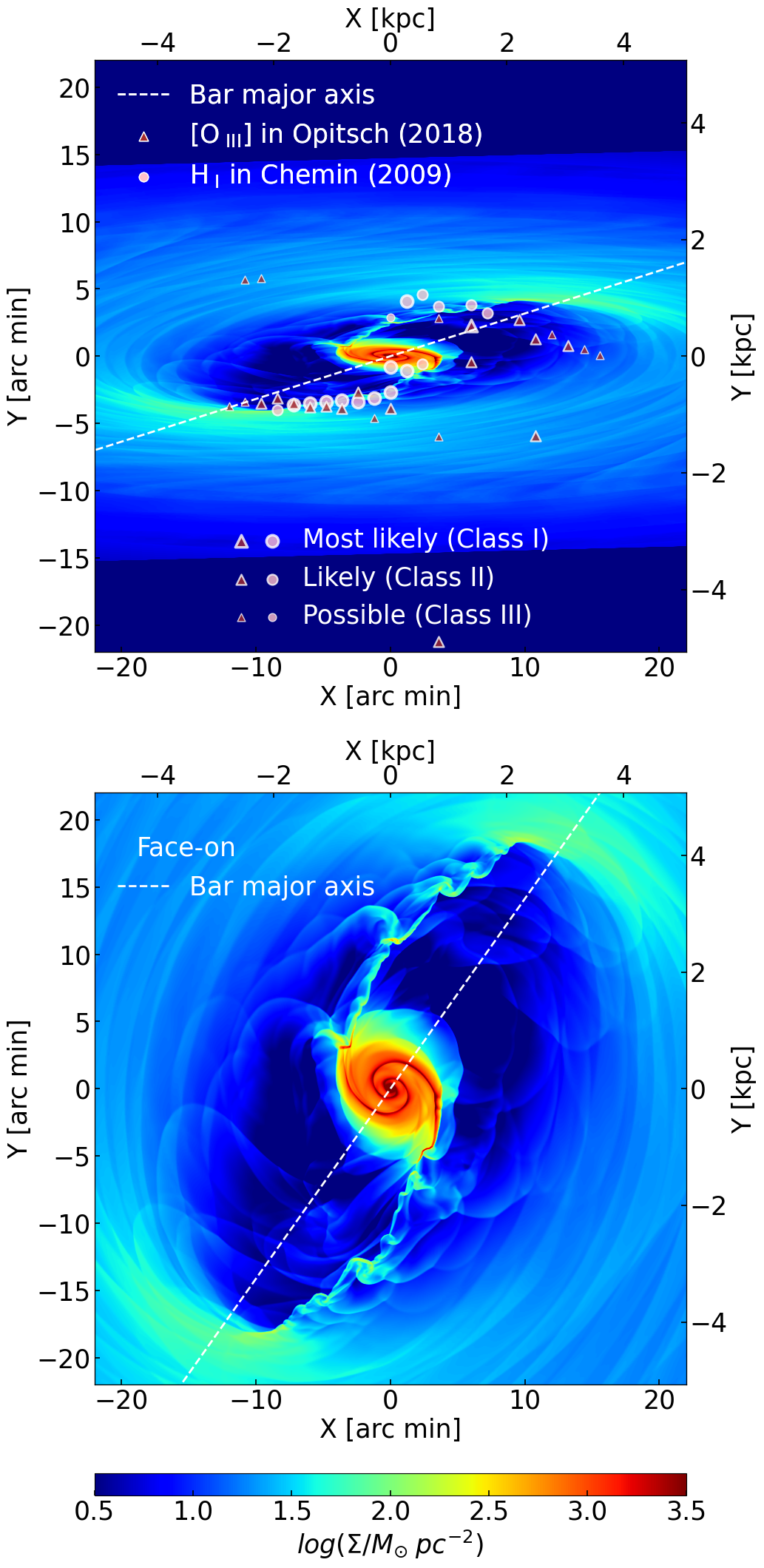}
\caption{\textit{Upper panel}: Gas surface density of Model 1 projected with an inclination of 77$^{\circ}$. The snapshot is at $T = 749 \Myr$. Color represents gas surface density in the model. Pink circles and brown triangles indicate the shock positions in $\firstcomponent$ and $\HI$, respectively. The orientation of the bar major axis in the model is shown by the white dashed line. \textit{Lower panel}: Face-on view of Model 1 at $T = 749 \Myr$. The disk major axis (inclination axis) is along the X-axis.}
\label{fig:shock_position_model1}
\end{figure}

\subsection{Gravitational potential}

The galactic potential is based on the best fit m2m $N$-body model constructed by \citet{blana_etal_18}. The m2m model of M31 \citep{blana_etal_18} was derived by fitting the IRAC-$3.6\;{\mu}m$ photometry \citep{bar_etal_06}, the IFU stellar kinematics in the bulge \citep{opitsc_etal_18} and the $\HI$ rotation curve \citep{cor_etal_10}. The triaxial bulge in the dynamical model consists of a classical bulge component with mass of $1.18^{+0.06}_{-0.07} \times 10^{10}\Msun$ and a BPB component with mass of $1.91 \;\pm\; 0.06 \times 10^{10}\Msun$. The bar in their model has a pattern speed of $\Omega_b = 40 \;\pm\; 5 \freq$ and a length of semi-major axis of 4 $\kpc$. The major axis of the bar is oriented at a position angle of 54.7$^{\circ}$ (in the galactic plane) with respect to the line of nodes. The dark matter halo in their model follows an Einasto profile and the mass of dark matter within the bulge region ($R < 3.2 \kpc$) is $1.2^{+0.2}_{-0.4} \times 10^{10}\Msun$. The authors obtained similar mass values using models with NFW dark matter profiles. The mass-to-light ratio of stellar component in $3.6\;{\mu}m$ is found to be $\Upsilon_{3.6} = 0.72\;\pm\; 0.02 \Msun\Lsun^{-1}$. We use their best fit models JR804 and KR241 as the basis of our galactic potential. The former includes an Einasto dark matter halo and the latter includes an NFW dark matter halo. We add a Plummer sphere in the center to represent the supermassive black hole with a mass of $\MBH = 2 \times 10^8\Msun$ \citep{ben_kor_05}:

\begin{equation}
\Phi(r) = -G\MBH\dfrac{1}{\sqrt{r^2+a^2}}.
\end{equation}

\noindent Here $a = 10\pc$.

\begin{figure}[t!]
\includegraphics[width=\columnwidth]{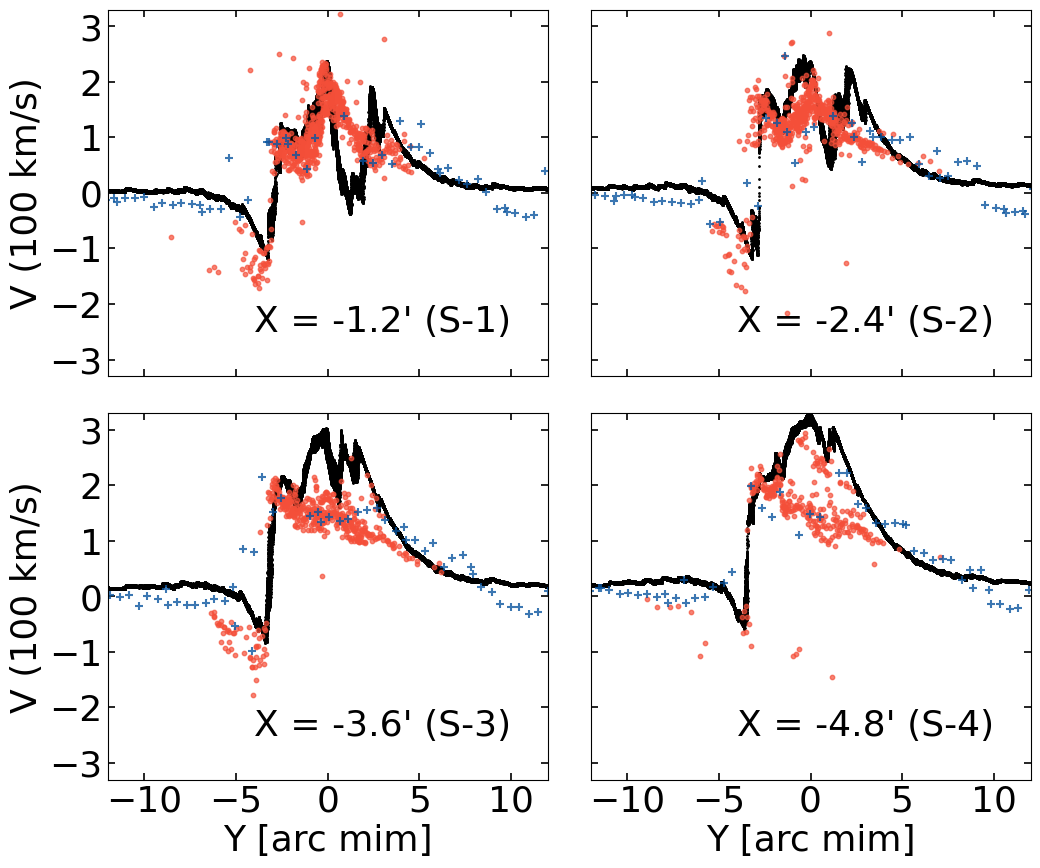}
\caption{Comparison between simulated and observed gas PVD for Model 1. Black dots represent the PVDs of the model along 4 pseudo-slits with their slit numbers shown on the top left. Red circles and blue crosses represent the observed PVDs of $\firstcomponent$ and main component of $\HI$, respectively.}
\label{fig:PVDs_model1}
\end{figure}

\begin{figure}[t!]
\includegraphics[width=\columnwidth]{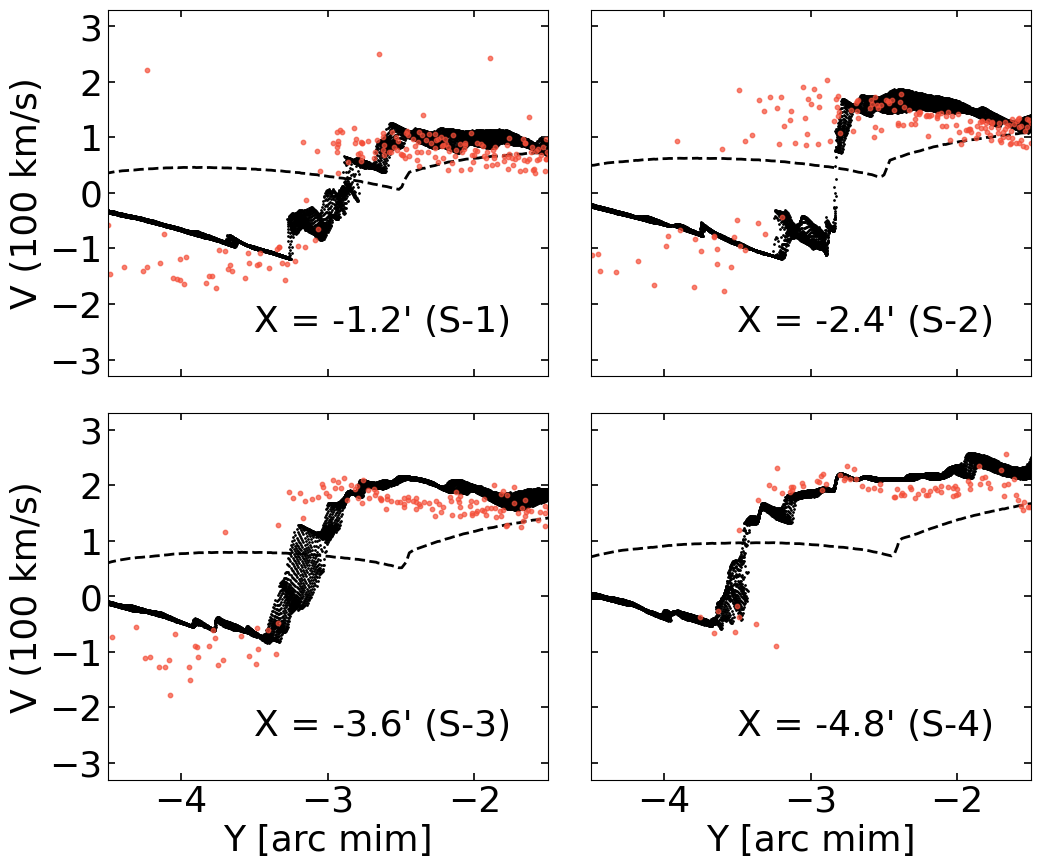}
\caption{Zoom-in version of Figure~\ref{fig:PVDs_model1}. We add black dashed lines to represent PVDs in a non-rotating bar model. The slit width for creating PVDs in this model is 0.04$\arcmin$ while for Model 1 it halves to 0.6$\arcmin$ (compared to the previous figure) for better clarity. The $\HI$ component is not shown in this figure.}
\label{fig:PVDs_model1_zoomin}
\end{figure}

\subsection{Simulated bar-driven shocks}

For simplicity, we only compare simulated bar-driven shocks with the clearest shock features of $\firstcomponent$ on the far side (slits S-1, S-2, S-3, S-4 in Fig.~\ref{fig:PVDs_component_1_receding_side}). We discuss the possible reasons for weaker shocks on the near side in \S\ref{sec:asytwoside}. We start with models using a bar pattern speed around $\Omega_b = 40 \freq$ given in \citet{blana_etal_18}. After projection with an inclination of $77^{\circ}$, the shocks are found to be too close to the bar major axis and cannot extend as far as shock positions in $\firstcomponent$. Other parameters that could be important are the uncertain shape and strength of the thin bar. \citet{ham_etal_18} suggested that a single merger event $\sim 2 \Gyr$ ago could explain the recent active star formation \citep{wil_etal_15}. More recent work studied the age-velocity dispersion relation in M31's stellar disk \citep{bha_etal_19}, which suggests a merger with a mass ratio of around 1:5 3-4 $\Gyr$ ago. The bar could have been weakened by the merger \citep{gho_etal_21} and left its shape and strength uncertain. One way to produce shocks with a larger distance from the bar major axis is to decrease the bar pattern speed \citep{li_etal_15}. We tested
models with lower bar pattern speeds and their shocks have a better match with $\firstcomponent$ data. When the bar rotates with a lower pattern speed around $\Omega_b = 20\freq$, the nuclear ring size becomes very large and the effects of thermal pressure need to be considered to reduce the size of the nuclear ring. It requires a high $c_s$ of around $30 \kms$ to make a reasonably sized nuclear ring. Models with low $\Omega_b$ and high $c_s$ produce less curved shocks, larger velocity jumps, and a nuclear ring with a reasonable size. Using an Einasto or NFW dark matter halo does not affect much the substructures.

We present the shock pattern in two models: (1) Model 1 with a bar pattern speed $\Omega_b = 20\freq$ and sound speed $c_s = 30\kms$ based on JR804 potential; (2) Model 2 with a bar pattern speed $\Omega_b = 33\freq$ and sound speed of $c_s = 15\kms$ based on KR241 potential. The bar rotation period $2\pi/{\Omega_{b}}$ for Model 1 and Model 2 is 307$\Myr$ and 186$\Myr$, respectively. The ratio of co-rotation radius to semi-major axis of the bar is $\mathcal{R}\equiv R_{CR}/a_{bar} \sim 3.2$ and $\mathcal{R} \sim 1.8$ for Model 1 and Model 2, respectively. Although simulations have reached quasi-steady after two bar rotation periods, small transient changes still appear on PVDs. We choose snapshots such that the shock features of the models are most similar to those of $\firstcomponent$.

The $x$-axis of the simulation grid is along the major axis of the disk. The major axis of the bar in Model 1 is positioned at an angle of 54.7$^{\circ}$ to the $x$-axis in the face-on case. The upper panel of Fig.~\ref{fig:shock_position_model1} illustrates the gas surface density of Model 1 at T = 749$\Myr$ projected with an inclination of 77$^{\circ}$. Pink circles and brown triangles represent the shock positions of $\oiii$ and $\HI$, respectively. Large, medium-sized, and small markers represent the Class I, Class II, and Class III shock features. We present the face-on view of Model 1 in the lower panel of Fig.~\ref{fig:shock_position_model1}. In Fig.~\ref{fig:PVDs_model1} we present the PVDs of Model 1 on the receding side. Red circles and blue plus signs represent the PVDs of $\oiii$ and the main component of $\HI$, respectively. The average shock velocity jump $\Delta V$ is around 310$\kms$ in both Model 1 and $\firstcomponent$. Fig.~\ref{fig:PVDs_model1} shows that the profiles on PVDs of Model 1, $\firstcomponent$ and $\HI$ are roughly consistent. In addition, we further show in Fig.~\ref{fig:PVDs_model1_zoomin} that a non-rotating bar (i.e. similar to a triaxial bulge) cannot reproduce the shock features. The black dashed lines in Fig.~\ref{fig:PVDs_model1_zoomin} represent the PVDs of a model that has the exact same setups with Model 1 but with zero bar pattern speed. It is clear that no sharp velocity jumps can be found in this non-rotating bar model.

Adjusting the inclination and the bar angle of models can fine-tune the match with the observed shock positions of $\firstcomponent$. As the inclination decreases, the angle of the bar (after projection) in respect of the line of nodes becomes larger, resulting in shock positions further away from the disk major axis. In the process of decreasing inclination, a smaller bar angle is needed to keep the shape of shocks. 
\begin{figure}[t!]
\includegraphics[width=\columnwidth]{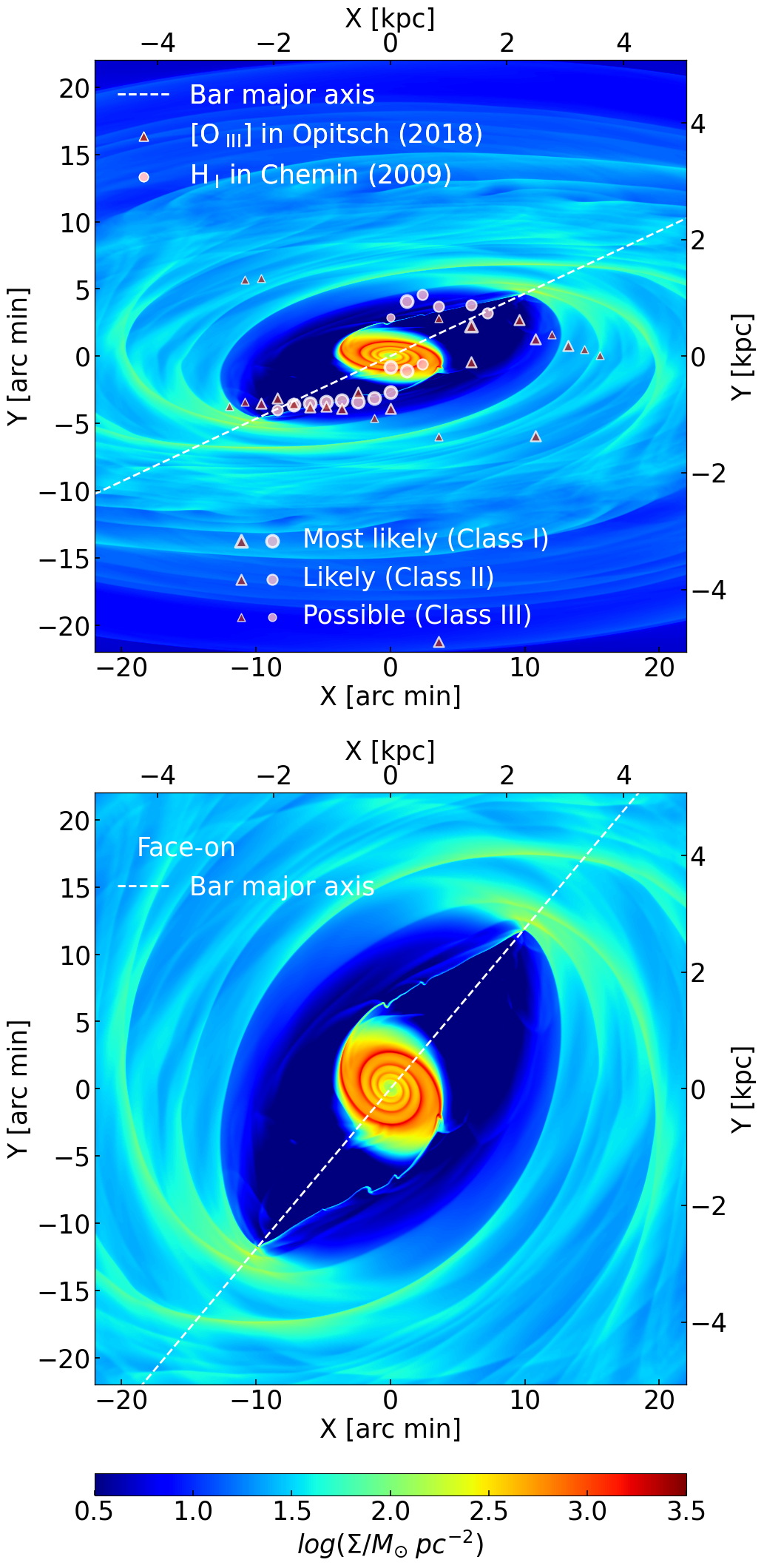}
\caption{\textit{Upper panel}: Gas surface density of Model 2 projected with an inclination of 67$^{\circ}$. The snapshot is at $T = 799 \Myr$. Color and markers have the same meaning as in Fig.~\ref{fig:shock_position_model1}. \textit{Lower panel}: Face-on view of Model 2 at $T = 799 \Myr$.}
\label{fig:shock_position_model2}
\end{figure}

\begin{figure}[t!]
\includegraphics[width=\columnwidth]{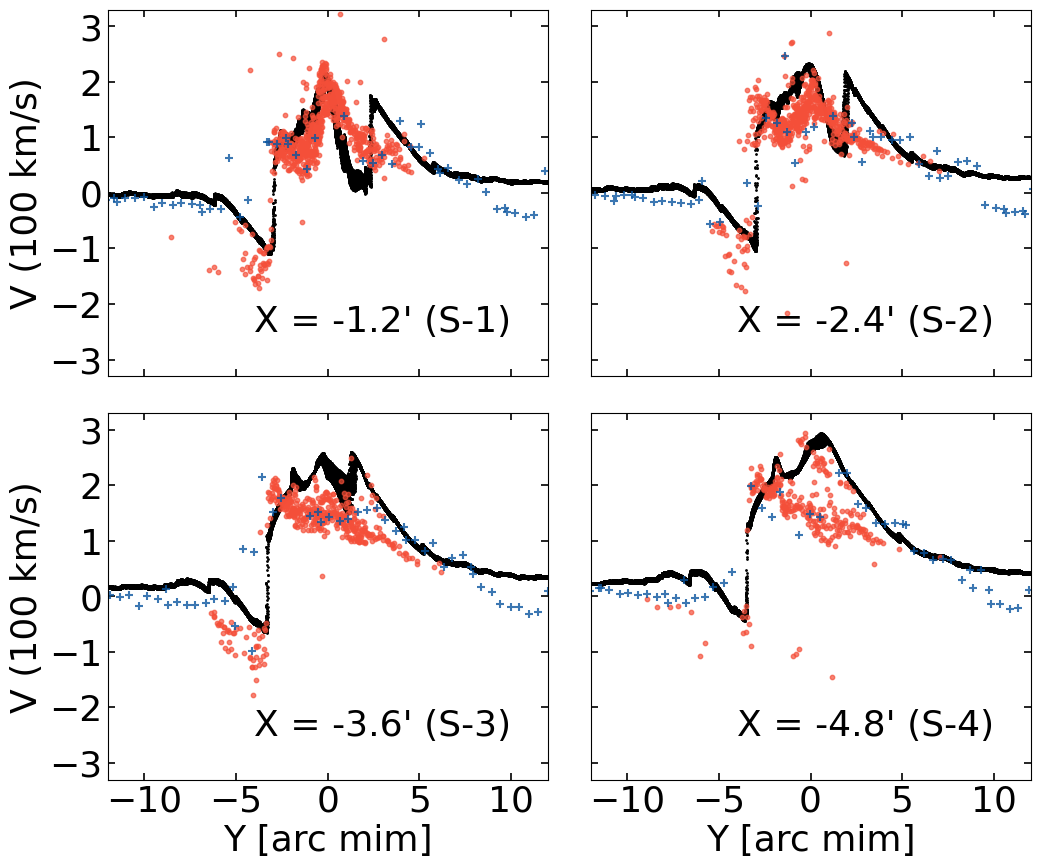}
\caption{Similar to Figure~\ref{fig:PVDs_model1} but for Model 2.}
\label{fig:PVDs_model2}
\end{figure}

\begin{figure}[t!]
\includegraphics[width=\columnwidth]{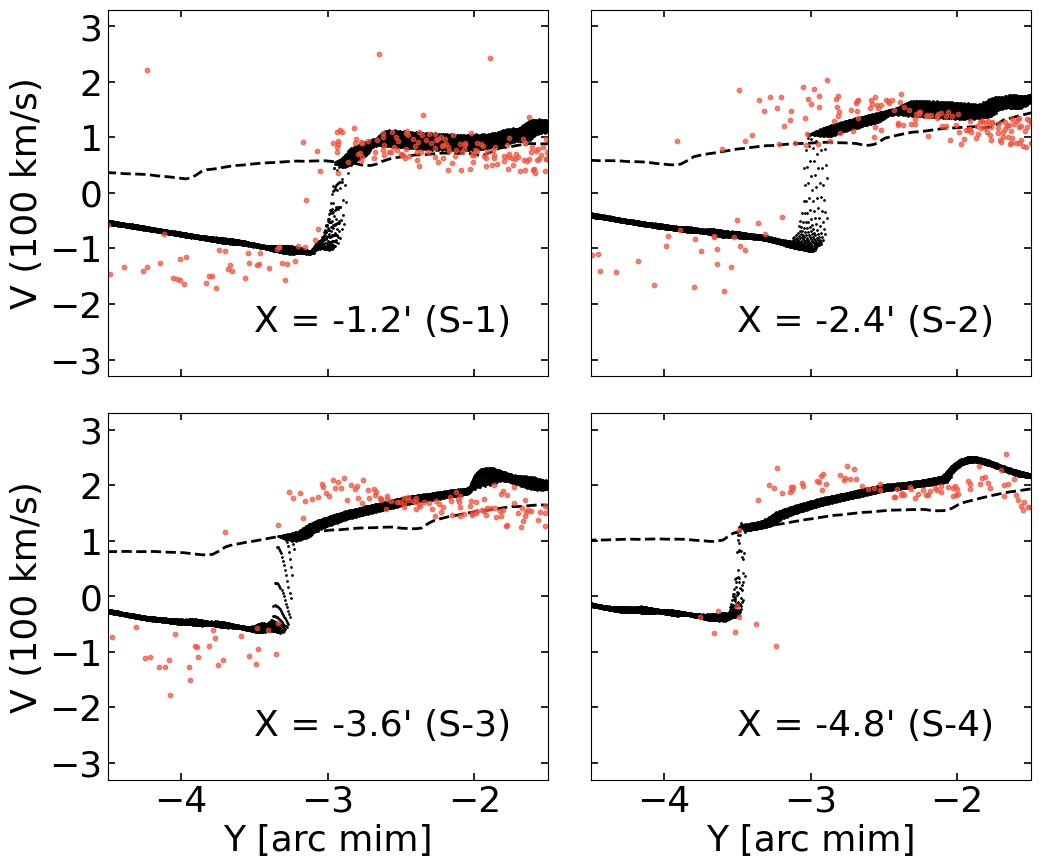}
\caption{Similar to Figure~\ref{fig:PVDs_model1_zoomin} but for Model 2 (black dots) and its corresponding non-rotating bar model (dashed lines)}
\label{fig:PVDs_model2_zoomin}
\end{figure}

We run Model 2 to test the effects of inclination and bar angle. Model 2 produces shocks much closer to the bar major axis than Model 1, which shows a different shock pattern from that in $\firstcomponent$. However, with a smaller inclination of 67$^{\circ}$ and a bar angle of 50$^{\circ}$, shock positions in Model 2 can still be similar to those in $\firstcomponent$, especially on the far side. In the upper panel of Fig.~\ref{fig:shock_position_model2} we present the gas surface density of Model 2 at T = 799$\Myr$ projected with an inclination of 67$^{\circ}$. The lower panel of Fig.~\ref{fig:shock_position_model2} shows the face-on view of Model 2. Fig.~\ref{fig:PVDs_model2} illustrates the PVDs of Model 2 on the receding side and its comparison with $\firstcomponent$ and $\HI$. The average shock velocity jump $\Delta V$ is smaller than $\firstcomponent$ by $\approx 30\%$. These results show that there may be a degeneracy between the inclination and the bar pattern speed. We also present the PVDs in a non-rotating Model 2 that has zero bar pattern speed in Fig.~\ref{fig:PVDs_model2_zoomin}. There are no shocks in this model either. Figs.~\ref{fig:PVDs_model1_zoomin} and \ref{fig:PVDs_model2_zoomin} demonstrate that bar rotation is necessary to produce the observed shock features.

Both Model 1 and Model 2 have shock positions roughly similar to $\firstcomponent$ on the far side. The main advantage of Model 1 is the large shock velocity jumps similar to those in $\firstcomponent$, but the bar parameters in Model 2 are closer to those obtained from the m2m models \citep[][]{blana_etal_18}. We also tested different pattern speeds within the range of 25 - 50 $\freq$ and sound speeds within the range of 1 - 50 $\kms$. Shock positions move closer to the $x$-axis with a larger bar pattern speed and/or a larger sound speed. The line-of-sight velocity jump at shocks becomes larger as bar pattern speed decreases and sound speed increases. A detailed comparison including more bar parameters will be presented in a follow-up paper.

\citet{berman_01} and \citet{ber_loi_02} performed hydrodynamical simulations in a M31 potential and found a much higher bar pattern speed of $\Omega_b = 53.7\kms$ by fitting the line-of-sight velocity of CO \citep{loinar_95,loinar_99} along the disk major axis. They adopted a simple analytical bar model, and the spatial resolution of their simulations was relatively low ($\sim 125 \pc$, in contrast to to $7.3 \pc$ in this work). Their model can match the PVDs of CO, which, unfortunately, do not cover the shock regions of $\oiii$ and $\HI$ used in this work. The dust lanes predicted by their models are nearly perpendicular to the disk major axis, and are quite different from the position of shock features in Fig.~\ref{fig:position_shock}.

\begin{figure}[t!]
\includegraphics[width=\columnwidth]{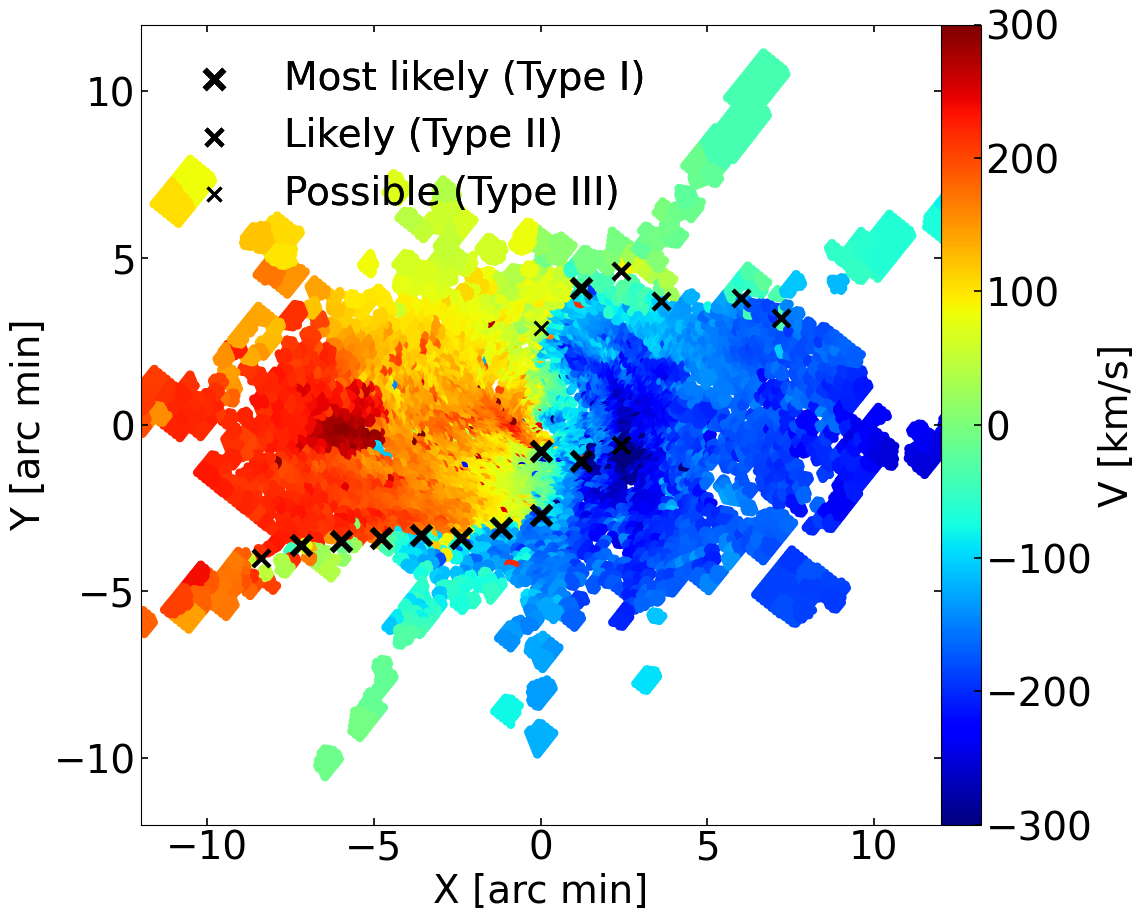}
\caption{Positions of shock features of $\firstcomponent$ overlaid on its velocity map. Large, medium-sized and small crosses represent the Class I, Class II, and Class III shock features. The velocity field of $\firstcomponent$ is asymmetric between the far (bottom left) and near (top right) side. A large velocity gradient is present near the zero-velocity line on the far side, which locates mostly within the boundary. Class I shock features closely follow the zero-velocity line. On the near side, the velocity field is smoother and we can only identify weak shock features near the boundary.}
\label{fig:velocity_map_first_component}
\end{figure}

\section{Discussion}

\subsection{Asymmetry of shocks between the near and far side}
\label{sec:asytwoside}

We find that the shock features are generally weaker on the near side than on the far side. In Fig.~\ref{fig:PVDs_component_1_approaching_side} it is clear that shock features are absent near $Y\sim5\arcmin$ in slit S+4. In Fig.~\ref{fig:velocity_map_first_component} we overlay the positions of the $\firstcomponent$ shocks on the observed 2D velocity map. Class I shock features (large crosses) are found mainly at $-7.2\arcmin < X < 0\arcmin$ on the far side (bottom left). On the near side (top right), there are mostly Class II shock features (medium-sized crosses) in regions with small velocity gradients. Overall shock features are closer to the boundary of data coverage on the near side than the far side. This implies the shock features of $\firstcomponent$ defined in the current study might be incomplete due to the limited coverage of observation.

Clear asymmetry in the position of shocks and gas kinematics has been observed in the barred galaxy NGC 1365 \citep{zan_etal_08}. They suggest that the asymmetry in positions of the dust lane is likely caused by a minor merger event. They also provided an alternative explanation that the ram pressure of a gas stream has moved the shock to offset its original position.

\begin{figure}[t!]
\includegraphics[width=\columnwidth]{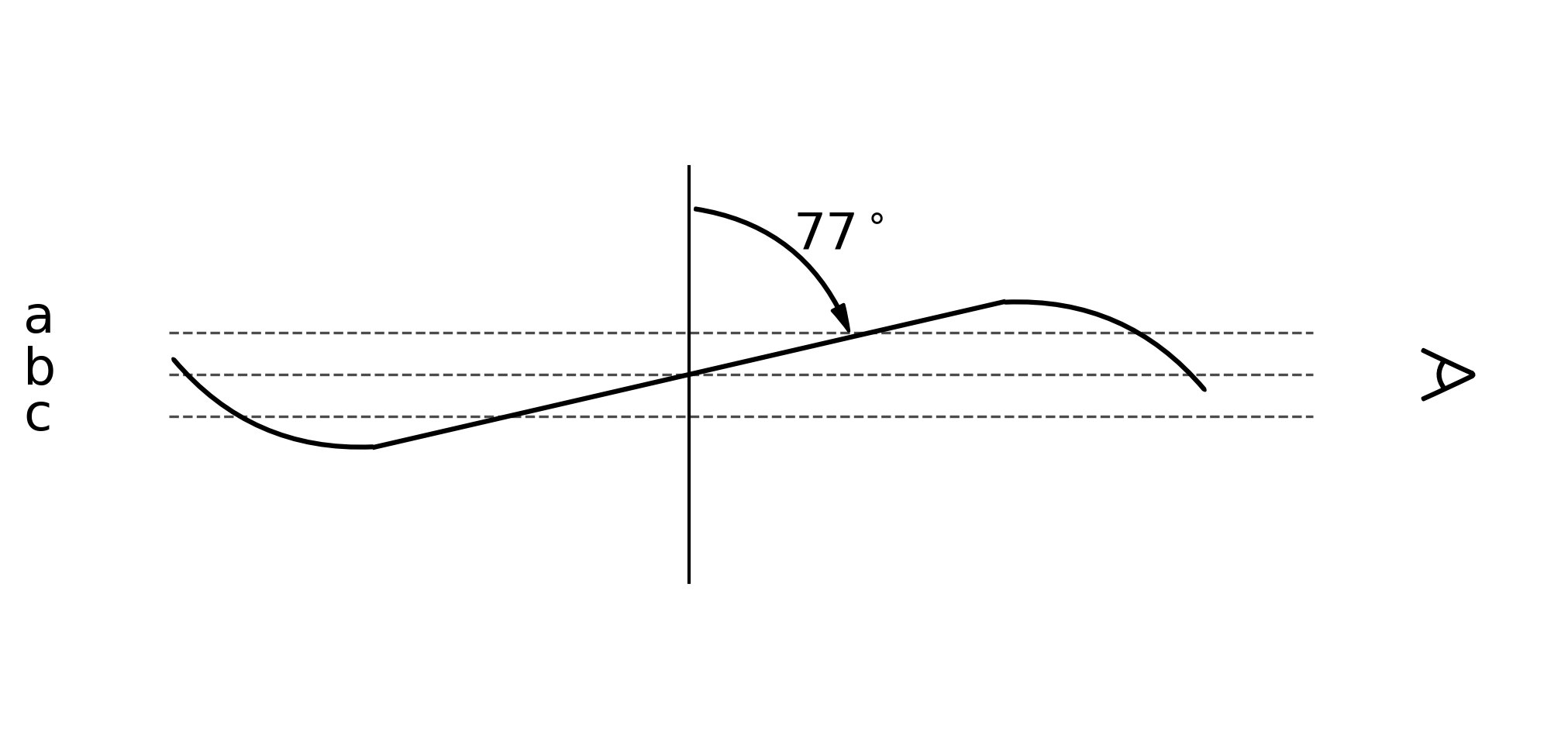}
\caption{Schematic view of the asymmetric extinction by the warp in M31's gas disk. Dotted lines indicate three possible lines of sight toward M31. (a) The line of sight crosses both the warp region and the inner disk on the near side. The gas and the dust on the warp become foreground and thus obscure the light from shock features. (b) The line of sight passes through the central regions of M31. Light is dominated by the outer warp for the absence of cold gas near the center. (c) The line of sight first crosses the inner disk, then passes the outer warp region. The warp becomes a background and it does not obscure the light from the inner disk.}
\label{fig:scheme_warp}
\end{figure}

Another possible scenario for the asymmetry in the shock feature is that the warp in the outer $\HI$ disk has a stronger extinction effect towards the near side. The $\HI$ disk is extended and has a large-scale warp in the outer region \citep{new_eme_77,hender_79,bri_bur_84,chemin_etal_09,cor_etal_10}. Fig.~\ref{fig:scheme_warp} presents a schematic diagram of the gas warp structure. The gas warp appears as a foreground on the near side and becomes a background on the far side. The asymmetric extinction by the gas warp could result in shock features different between the near side and the far side. Furthermore, abundant dust in the $\HI$ warp may also help blur the $\firstcomponent$ shock features on the near side. The distribution of dust is found to follow that of $\HI$ in the outer disk, and the strong reddening effect suggests a large amount of dust \citep{cui_etal_01, ber_etal_12}. More recent detection of dust by \citet{ruo_hai_20} found that dust in the disk has an exponential distribution and extends over 2.5 times its optical radius (around 54 kpc).

\begin{figure}[!t]
\includegraphics[width=\columnwidth]{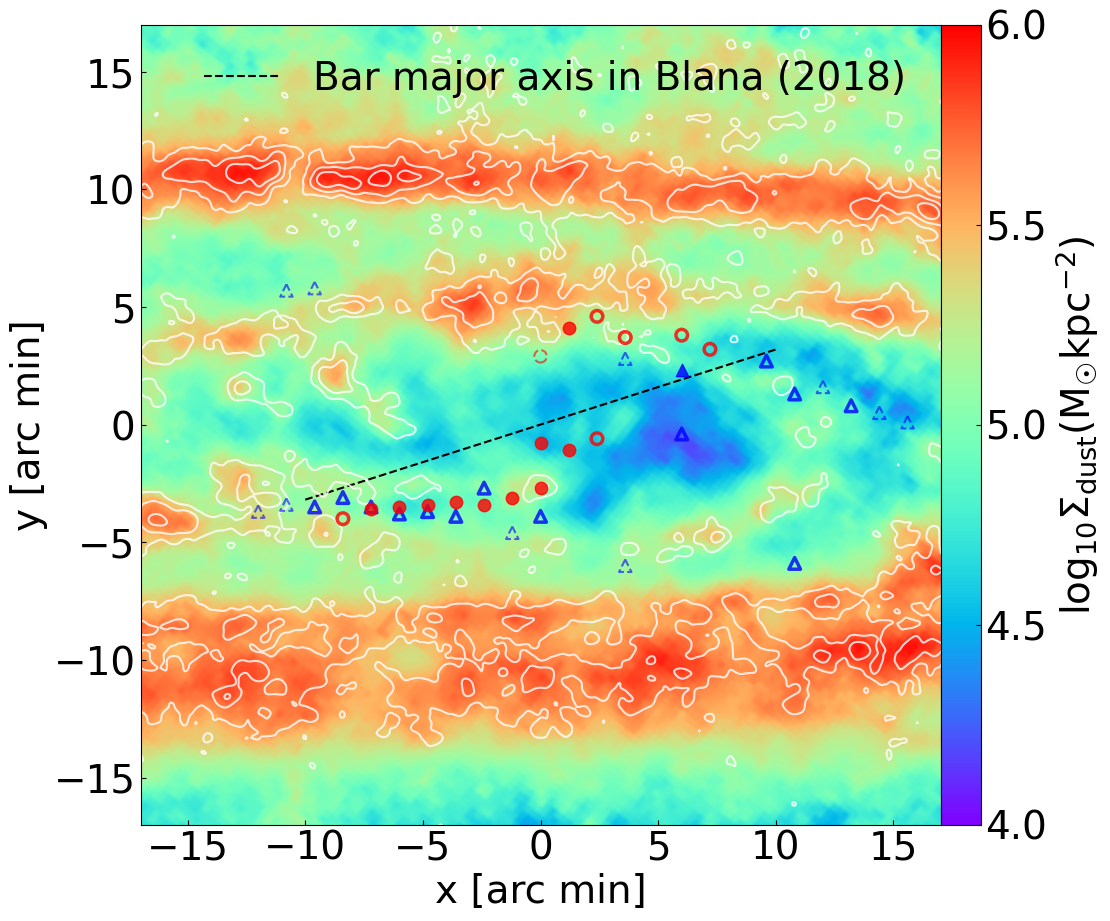} 
\caption{Identified shock positions of $\firstcomponent$ (red circles) and $\HI$ (blue triangles) superposed on the surface density map of dust \citep{dra_etal_14} and contours of CO emission \citep{nieten_etal_06}. The solid, open, and dashed markers represent the Class I, Class II, and Class III shock features, respectively. The contour levels are 1, 4, and 10 $K \kms$. The dashed line indicates the bar major axis in \citet{blana_etal_18}.}
\label{fig:compare_dust_CO}
\end{figure}

\subsection{Effects of changing slit orientations on shock features}

Maximum velocity jumps may be expected when slits are positioned perpendicular to the shock fronts \citep{athana_92}. We checked PVDs of $\firstcomponent$ and $\HI$ in slits at different orientations and compared them to Fig.~\ref{fig:scheme_shock} to see if the velocity jump features can be shown even more clearly. Our findings are roughly consistent with the theoretical expectation. When the slits are positioned parallel to the disk major axis, the shock features still appear, but with a smoother profile. The shock features become sharper as we reposition slits to be nearly perpendicular to the shock fronts, with an increase of $\Delta V$ around $30 \kms$ compared to the more parallel slits. In appendix \ref{appendix:B} we tested repositioning the slit to make it more perpendicular to the shock fronts and found that the shock properties are similar to \S\ref{sec:shock_features_first_component}.

\subsection{Comparison of shock features with dust and CO morphology}

In Fig.~\ref{fig:compare_dust_CO} we present the map of our identified shock features in $\firstcomponent$ (red circles) and $\HI$ (blue triangles) on a background of dust (color-coded with surface density) and CO (white contours) morphology. We use data of dust from \citet{dra_etal_14} and CO from \citet{nieten_etal_06}. The black solid line and dashed line indicate $PA_{disk}$ and the major axis of the bar in \citet{blana_etal_18}, respectively. The thickest dust arms correspond to the 10 $\kpc$ ring in M31. Positions of $\oiii$ and $\HI$ shock features are found near a thin dust lane on the far side. Such coincidence also appears in the CO distribution, which generally matches the dust morphology. The shock features of $\HI$, extending to the bar ends in \citet{blana_etal_18}, connect to a possible spiral structure. For the near side, shock features are weaker as we discussed in \S\ref{sec:asytwoside}. There are not enough Class II shock features to show the relation between shock positions and dust morphology. Another reason might be that the amount of $\HI$, CO, and dust in the bulge is scarce on the near side. Near center there is a lack of $\HI$ shock features and CO emission. This could be due to the low gas density in the central region \citep{bri_sha_84,li_etal_09,don_etal_16,li_etal_19,li_etal_20}.

\subsection{Comparison of shock features in other nearby barred galaxies}

Other barred galaxies also show qualitatively similar bar-induced shock features as in M31. \citet{mun_etal_99a} detected velocity jumps of $\HI$ with $\Delta V \approx 130 \kms$ on the leading side of bar of NGC 4151. $\HI$ in NGC 4151 is not interfered by the gas warp, showing a clearer view of shock features. The $\rm H\alpha$ velocity field of the barred galaxy NGC 4123 illustrates similar shock features with $\Delta V > 100\kms$ \citep{wei_etal_01a}. Velocity jumps in NGC 4151 and NGC 4123 have a smaller amplitude than our Class I shock features, probably due to the lower inclinations and lower masses of galaxies. A large velocity gradient near dust lanes has been observed in other barred galaxies as well, e.g. NGC 1530 \citep{rey_dow_98,zur_etal_04}, NGC 7479 \citep{lai_etal_1999}, NGC 5448 \citep{fat_etal_05}, NGC 1365 \citep{zan_etal_08}. The recent high-resolution PHANGS–ALMA survey presents 2D gas kinematics of CO for nearby spiral galaxies \citep{ler_etal_21}. The velocity gradient near dust lanes can even be visually identified in many barred galaxies from their sample, e.g. NGC 2903, NGC 3627, NGC 4536 and NGC 4945. Apart from bar-driven shocks, spiral arms in non-barred galaxies could also produce velocity jumps, but with a much smaller amplitude \citep[usually around 40 $\kms$ in simulations][]{robert_69,pet_etal_20}.

\subsection{Origin of the two velocity components}
\label{sec:twooxy3}

We use $\firstcomponent$ to present PVDs in the major part of the paper, but there is another component $\secondcomponent$ observed in \citet{opitsc_etal_18}. The gas flow forms streams and ring structures in barred potentials \citep{kim_etal_12a}. Multiple gas components could be found when the line of sight passes through several of such gas substructures. A collision between M31 and its satellite galaxy M32 \citep{block_etal_06} could also produce gas streams and rings, leading to multiple observed gas components along the line of sight. It is also possible that part of $\oiii$ comes from the bulge instead of the disk in M31. 

Considering that the $\secondcomponent$ may be a foreground or background, we do not expect the $\secondcomponent$ to show clear shock features. It is interesting that PVDs of $\secondcomponent$ show a few velocity jump features, but the amplitude is generally small and their positions do not show a regular pattern.

Apart from the ionized gas in the optical band, CO and $\HI$ observations also revealed multiple gas spectral lines \citep{mel_com_11,chemin_etal_09}. \citet{mel_com_11} proposed a model of the tilted rings to explain the multiple lines in CO. However, it seems that their scenario cannot explain the shape of $\oiii$ spectra well. \citet{chemin_etal_09} attributed the lower-velocity gas components to the outer $\HI$ warp, but whether the warp may cause a split $\oiii$ remains unclear. Future studies are needed to provide a better explanation for the $\secondcomponent$ component.  

\section{Conclusion}

We identify shock features in the central region of M31 ($20\arcmin \times 10\arcmin$) using $\oiii$ data from \citet{opitsc_etal_18} and $\HI$ data from \citet{chemin_etal_09}.  The strongest $\oiii$ shock features show a large velocity gradient (over 1.2 $\kmspc$) with $\Delta V$ over 170 $\kms$ in the bulge. The $\HI$ emission shows similar shock features even beyond the bulge region. Note that several shock features show up in $\oiii$ but not in $\HI$ emission near the center, possibly due to the lack of atomic gas there. The shock features are found mainly on the leading side of the possible bar proposed by \citet{blana_etal_18}.
The spatial location of the shocks and the amplitude of shock velocity jumps are qualitatively consistent with our preliminary simulations of bar-induced gas inflow in M31. This result provides independent, strong evidence that M31 hosts a large bar. A detailed comparison with more hydrodynamical simulations will be presented in a follow-up study to provide a better understanding of the gas features in the center of M31, and hopefully determine better the main bar parameters of M31.

\software{
{\tt Athena++} \citep{sto_etal_20},
NumPy \citep{2020NumPy-Array},
SciPy \citep{2020SciPy-NMeth},
Matplotlib \citep{4160265},
Jupyter Notebook \citep{Kluyver2016jupyter}
}

\begin{acknowledgments}

The authors would like to thank Laurent Chemin for sharing $\HI$ data of M31 and the anonymous referee for an encouraging and constructive report. The research presented here is partially supported by the National Key R\&D Program of China under grant No. 2018YFA0404501; by the National Natural Science Foundation of China under grant Nos. 12103032, 12025302, 11773052, 11761131016; by the ``111'' Project of the Ministry of Education of China under grant No. B20019; and by the China Manned Space Project under grant No. CMS-CSST-2021-B03.  J.S. also acknowledges support from a \textit{Newton Advanced Fellowship} awarded by the Royal Society and the Newton Fund. M.B. acknowledges funding from ANID through the FONDECYT Postdoctorado 2021 Nr 3210592 and the Excellence Cluster ORIGINS founded by the Deutsche Forschungsgemeinschaft (DFG; German Research Foundation) under Germany’s Excellence Strategy – EXC-2094 – 390783311, the supports from BASAL CATA Center for Astrophysics and Associated Technologies, and the Max Planck Computing and Data Facility. This work made use of the Gravity Supercomputer at the Department of Astronomy, Shanghai Jiao Tong University, and the facilities of the Center for High Performance Computing at Shanghai Astronomical Observatory. 
\end{acknowledgments}

\bibliographystyle{apj}
\bibliography{ms}

\begin{thebibliography}{}
\expandafter\ifx\csname natexlab\endcsname\relax\def\natexlab#1{#1}\fi

\bibitem[{{Athanassoula}(1992)}]{athana_92}
{Athanassoula}, E. 1992, \mnras, 259, 345

\bibitem[{{Athanassoula} \& {Beaton}(2006)}]{ath_bea_06}
{Athanassoula}, E., \& {Beaton}, R.~L. 2006, \mnras, 370, 1499

\bibitem[{{Athanassoula} \& {Bureau}(1999)}]{ath_bur_99}
{Athanassoula}, E., \& {Bureau}, M. 1999, \apj, 522, 699

\bibitem[{{Barmby} {et~al.}(2006){Barmby}, {Ashby}, {Bianchi}, {Engelbracht},
  {Gehrz}, {Gordon}, {Hinz}, {Huchra}, {Humphreys}, {Pahre},
  {P{\'e}rez-Gonz{\'a}lez}, {Polomski}, {Rieke}, {Thilker}, {Willner}, \&
  {Woodward}}]{bar_etal_06}
{Barmby}, P., {Ashby}, M.~L.~N., {Bianchi}, L., {et~al.} 2006, \apjl, 650, L45

\bibitem[{{Beaton} {et~al.}(2007){Beaton}, {Majewski}, {Guhathakurta},
  {Skrutskie}, {Cutri}, {Good}, {Patterson}, {Athanassoula}, \&
  {Bureau}}]{beaton_etal_07}
{Beaton}, R.~L., {Majewski}, S.~R., {Guhathakurta}, P., {et~al.} 2007, \apjl,
  658, L91

\bibitem[{{Bender} {et~al.}(2005){Bender}, {Kormendy}, {Bower}, {Green},
  {Thomas}, {Danks}, {Gull}, {Hutchings}, {Joseph}, {Kaiser}, {Lauer},
  {Nelson}, {Richstone}, {Weistrop}, \& {Woodgate}}]{ben_kor_05}
{Bender}, R., {Kormendy}, J., {Bower}, G., {et~al.} 2005, \apj, 631, 280

\bibitem[{{Berman}(2001)}]{berman_01}
{Berman}, S. 2001, \aap, 371, 476

\bibitem[{{Berman} \& {Loinard}(2002)}]{ber_loi_02}
{Berman}, S., \& {Loinard}, L. 2002, \mnras, 336, 477

\bibitem[{{Bernard} {et~al.}(2012){Bernard}, {Ferguson}, {Barker}, {Hidalgo},
  {Ibata}, {Irwin}, {Lewis}, {McConnachie}, {Monelli}, \&
  {Chapman}}]{ber_etal_12}
{Bernard}, E.~J., {Ferguson}, A. M.~N., {Barker}, M.~K., {et~al.} 2012, \mnras,
  420, 2625

\bibitem[{{Bertola} {et~al.}(1988){Bertola}, {Vietri}, \&
  {Zeilinger}}]{ber_etal_88}
{Bertola}, F., {Vietri}, M., \& {Zeilinger}, W.~W. 1988, The Messenger, 52, 24

\bibitem[{{Bhattacharya} {et~al.}(2019){Bhattacharya}, {Arnaboldi}, {Caldwell},
  {Gerhard}, {Bla{\~n}a}, {McConnachie}, {Hartke}, {Guhathakurta}, {Pulsoni},
  \& {Freeman}}]{bha_etal_19}
{Bhattacharya}, S., {Arnaboldi}, M., {Caldwell}, N., {et~al.} 2019, \aap, 631,
  A56

\bibitem[{{Bla{\~n}a D{\'\i}az} {et~al.}(2017){Bla{\~n}a D{\'\i}az}, {Wegg},
  {Gerhard}, {Erwin}, {Portail}, {Opitsch}, {Saglia}, \&
  {Bender}}]{bla_etal_17}
{Bla{\~n}a D{\'\i}az}, M., {Wegg}, C., {Gerhard}, O., {et~al.} 2017, \mnras,
  466, 4279

\bibitem[{{Bla{\~n}a D{\'\i}az} {et~al.}(2018){Bla{\~n}a D{\'\i}az}, {Gerhard},
  {Wegg}, {Portail}, {Opitsch}, {Saglia}, {Fabricius}, {Erwin}, \&
  {Bender}}]{blana_etal_18}
{Bla{\~n}a D{\'\i}az}, M., {Gerhard}, O., {Wegg}, C., {et~al.} 2018, \mnras,
  481, 3210

\bibitem[{{Block} {et~al.}(2006){Block}, {Bournaud}, {Combes}, {Groess},
  {Barmby}, {Ashby}, {Fazio}, {Pahre}, \& {Willner}}]{block_etal_06}
{Block}, D.~L., {Bournaud}, F., {Combes}, F., {et~al.} 2006, \nat, 443, 832

\bibitem[{{Brinks} \& {Burton}(1984)}]{bri_bur_84}
{Brinks}, E., \& {Burton}, W.~B. 1984, \aap, 141, 195

\bibitem[{{Brinks} \& {Shane}(1984)}]{bri_sha_84}
{Brinks}, E., \& {Shane}, W.~W. 1984, \aaps, 55, 179

\bibitem[{{Bureau} \& {Athanassoula}(1999)}]{bur_ath_99}
{Bureau}, M., \& {Athanassoula}, E. 1999, \apj, 522, 686

\bibitem[{Canny(1986)}]{canny_86}
Canny, J. 1986, IEEE Transactions on Pattern Analysis and Machine Intelligence,
  PAMI-8, 679

\bibitem[{{Chemin} {et~al.}(2009){Chemin}, {Carignan}, \&
  {Foster}}]{chemin_etal_09}
{Chemin}, L., {Carignan}, C., \& {Foster}, T. 2009, \apj, 705, 1395

\bibitem[{{Corbelli} {et~al.}(2010){Corbelli}, {Lorenzoni}, {Walterbos},
  {Braun}, \& {Thilker}}]{cor_etal_10}
{Corbelli}, E., {Lorenzoni}, S., {Walterbos}, R., {Braun}, R., \& {Thilker}, D.
  2010, \aap, 511, A89

\bibitem[{{Costantin} {et~al.}(2018){Costantin}, {M{\'e}ndez-Abreu}, {Corsini},
  {Eliche-Moral}, {Tapia}, {Morelli}, {Dalla Bont{\`a}}, \&
  {Pizzella}}]{cos_etal_18}
{Costantin}, L., {M{\'e}ndez-Abreu}, J., {Corsini}, E.~M., {et~al.} 2018, \aap,
  609, A132

\bibitem[{{Cuillandre} {et~al.}(2001){Cuillandre}, {Lequeux}, {Allen},
  {Mellier}, \& {Bertin}}]{cui_etal_01}
{Cuillandre}, J.-C., {Lequeux}, J., {Allen}, R.~J., {Mellier}, Y., \& {Bertin},
  E. 2001, \apj, 554, 190

\bibitem[{{Dong} {et~al.}(2016){Dong}, {Li}, {Wang}, {Lauer}, {Olsen}, {Saha},
  {Dalcanton}, \& {Groves}}]{don_etal_16}
{Dong}, H., {Li}, Z., {Wang}, Q.~D., {et~al.} 2016, \mnras, 459, 2262

\bibitem[{{Draine} {et~al.}(2014){Draine}, {Aniano}, {Krause}, {Groves},
  {Sandstrom}, {Braun}, {Leroy}, {Klaas}, {Linz}, {Rix}, {Schinnerer},
  {Schmiedeke}, \& {Walter}}]{dra_etal_14}
{Draine}, B.~T., {Aniano}, G., {Krause}, O., {et~al.} 2014, \apj, 780, 172

\bibitem[{{Emsellem} {et~al.}(2006){Emsellem}, {Fathi}, {Wozniak}, {Ferruit},
  {Mundell}, \& {Schinnerer}}]{emsell_06}
{Emsellem}, E., {Fathi}, K., {Wozniak}, H., {et~al.} 2006, \mnras, 365, 367

\bibitem[{{Fathi} {et~al.}(2005){Fathi}, {van de Ven}, {Peletier}, {Emsellem},
  {Falc{\'o}n-Barroso}, {Cappellari}, \& {de Zeeuw}}]{fat_etal_05}
{Fathi}, K., {van de Ven}, G., {Peletier}, R.~F., {et~al.} 2005, \mnras, 364,
  773

\bibitem[{{Gajda} {et~al.}(2021){Gajda}, {Gerhard}, {Bla{\~n}a}, {Zhu}, {Shen},
  {Saglia}, \& {Bender}}]{gaj_etal_21}
{Gajda}, G., {Gerhard}, O., {Bla{\~n}a}, M., {et~al.} 2021, \aap, 647, A131

\bibitem[{{Gerhard} {et~al.}(1989){Gerhard}, {Vietri}, \& {Kent}}]{ger_etal_89}
{Gerhard}, O.~E., {Vietri}, M., \& {Kent}, S.~M. 1989, \apjl, 345, L33

\bibitem[{{Ghosh} {et~al.}(2021){Ghosh}, {Saha}, {Di Matteo}, \&
  {Combes}}]{gho_etal_21}
{Ghosh}, S., {Saha}, K., {Di Matteo}, P., \& {Combes}, F. 2021, \mnras, 502,
  3085

\bibitem[{{Hammer} {et~al.}(2018){Hammer}, {Yang}, {Wang}, {Ibata}, {Flores},
  \& {Puech}}]{ham_etal_18}
{Hammer}, F., {Yang}, Y.~B., {Wang}, J.~L., {et~al.} 2018, \mnras, 475, 2754

\bibitem[{Harris {et~al.}(2020)Harris, Millman, van~der Walt, Gommers,
  Virtanen, Cournapeau, Wieser, Taylor, Berg, Smith, Kern, Picus, Hoyer, van
  Kerkwijk, Brett, Haldane, Fernández~del Río, Wiebe, Peterson,
  Gérard-Marchant, Sheppard, Reddy, Weckesser, Abbasi, Gohlke, \&
  Oliphant}]{2020NumPy-Array}
Harris, C.~R., Millman, K.~J., van~der Walt, S.~J., {et~al.} 2020, Nature, 585,
  357–362

\bibitem[{{Henderson}(1979)}]{hender_79}
{Henderson}, A.~P. 1979, \aap, 75, 311

\bibitem[{{Hunter}(2007)}]{4160265}
{Hunter}, J.~D. 2007, Computing in Science Engineering, 9, 90

\bibitem[{{Jacoby} {et~al.}(1985){Jacoby}, {Ford}, \&
  {Ciardullo}}]{jacoby_etal_85}
{Jacoby}, G.~H., {Ford}, H., \& {Ciardullo}, R. 1985, \apj, 290, 136

\bibitem[{{Kim} {et~al.}(2012){Kim}, {Seo}, {Stone}, {Yoon}, \&
  {Teuben}}]{kim_etal_12a}
{Kim}, W.-T., {Seo}, W.-Y., {Stone}, J.~M., {Yoon}, D., \& {Teuben}, P.~J.
  2012, \apj, 747, 60

\bibitem[{Kluyver {et~al.}(2016)Kluyver, Ragan-Kelley, P{\'e}rez, Granger,
  Bussonnier, Frederic, Kelley, Hamrick, Grout, Corlay, Ivanov, Avila, Abdalla,
  \& Willing}]{Kluyver2016jupyter}
Kluyver, T., Ragan-Kelley, B., P{\'e}rez, F., {et~al.} 2016, in Positioning and
  Power in Academic Publishing: Players, Agents and Agendas, ed. F.~Loizides \&
  B.~Schmidt, IOS Press, 87 -- 90

\bibitem[{{Kuzio de Naray} {et~al.}(2009){Kuzio de Naray}, {Zagursky}, \&
  {McGaugh}}]{kuzio_etal_09}
{Kuzio de Naray}, R., {Zagursky}, M.~J., \& {McGaugh}, S.~S. 2009, \aj, 138,
  1082

\bibitem[{{Laine} {et~al.}(1999){Laine}, {Kenney}, {Yun}, \&
  {Gottesman}}]{lai_etal_1999}
{Laine}, S., {Kenney}, J.~D.~P., {Yun}, M.~S., \& {Gottesman}, S.~T. 1999,
  \apj, 511, 709

\bibitem[{{Leroy} {et~al.}(2021){Leroy}, {Schinnerer}, {Hughes}, {Rosolowsky},
  {Pety}, {Schruba}, {Usero}, {Blanc}, {Chevance}, {Emsellem}, {Faesi},
  {Herrera}, {Liu}, {Meidt}, {Querejeta}, {Saito}, {Sandstrom}, {Sun},
  {Williams}, {Anand}, {Barnes}, {Behrens}, {Belfiore}, {Benincasa},
  {Be{\v{s}}li{\'c}}, {Bigiel}, {Bolatto}, {den Brok}, {Cao}, {Chandar},
  {Chastenet}, {Chiang}, {Congiu}, {Dale}, {Deger}, {Eibensteiner}, {Egorov},
  {Garc{\'\i}a-Rodr{\'\i}guez}, {Glover}, {Grasha}, {Henshaw}, {Ho}, {Kepley},
  {Kim}, {Klessen}, {Kreckel}, {Koch}, {Kruijssen}, {Larson}, {Lee}, {Lopez},
  {Machado}, {Mayker}, {McElroy}, {Murphy}, {Ostriker}, {Pan}, {Pessa},
  {Puschnig}, {Razza}, {S{\'a}nchez-Bl{\'a}zquez}, {Santoro}, {Sardone},
  {Scheuermann}, {Sliwa}, {Sormani}, {Stuber}, {Thilker}, {Turner}, {Utomo},
  {Watkins}, \& {Whitmore}}]{ler_etal_21}
{Leroy}, A.~K., {Schinnerer}, E., {Hughes}, A., {et~al.} 2021, \apjs, 257, 43

\bibitem[{{Li} {et~al.}(2016){Li}, {Gerhard}, {Shen}, {Portail}, \&
  {Wegg}}]{li_etal_16}
{Li}, Z., {Gerhard}, O., {Shen}, J., {Portail}, M., \& {Wegg}, C. 2016, \apj,
  824, 13

\bibitem[{{Li} {et~al.}(2020){Li}, {Li}, {Smith}, \& {Gao}}]{li_etal_20}
{Li}, Z., {Li}, Z., {Smith}, M. W.~L., \& {Gao}, Y. 2020, \apj, 905, 138

\bibitem[{{Li} {et~al.}(2019){Li}, {Li}, {Zhou}, {Gao}, {Jiang}, \&
  {Dong}}]{li_etal_19}
{Li}, Z., {Li}, Z., {Zhou}, P., {et~al.} 2019, \mnras, 484, 964

\bibitem[{{Li} {et~al.}(2015){Li}, {Shen}, \& {Kim}}]{li_etal_15}
{Li}, Z., {Shen}, J., \& {Kim}, W.-T. 2015, \apj, 806, 150

\bibitem[{{Li} {et~al.}(2009){Li}, {Wang}, \& {Wakker}}]{li_etal_09}
{Li}, Z., {Wang}, Q.~D., \& {Wakker}, B.~P. 2009, \mnras, 397, 148

\bibitem[{{Lindblad}(1956)}]{lindbl_56}
{Lindblad}, B. 1956, Stockholms Observatoriums Annaler, 19, 2

\bibitem[{{Loinard} {et~al.}(1995){Loinard}, {Allen}, \& {Lequeux}}]{loinar_95}
{Loinard}, L., {Allen}, R.~J., \& {Lequeux}, J. 1995, \aap, 301, 68

\bibitem[{{Loinard} {et~al.}(1999){Loinard}, {Dame}, {Heyer}, {Lequeux}, \&
  {Thaddeus}}]{loinar_99}
{Loinard}, L., {Dame}, T.~M., {Heyer}, M.~H., {Lequeux}, J., \& {Thaddeus}, P.
  1999, \aap, 351, 1087

\bibitem[{{McConnachie} {et~al.}(2005){McConnachie}, {Irwin}, {Ferguson},
  {Ibata}, {Lewis}, \& {Tanvir}}]{mcc_etal_05}
{McConnachie}, A.~W., {Irwin}, M.~J., {Ferguson}, A.~M.~N., {et~al.} 2005,
  \mnras, 356, 979

\bibitem[{{Melchior} \& {Combes}(2011)}]{mel_com_11}
{Melchior}, A.~L., \& {Combes}, F. 2011, \aap, 536, A52

\bibitem[{{M{\'e}ndez-Abreu} {et~al.}(2010){M{\'e}ndez-Abreu}, {Simonneau},
  {Aguerri}, \& {Corsini}}]{men_etal_10}
{M{\'e}ndez-Abreu}, J., {Simonneau}, E., {Aguerri}, J.~A.~L., \& {Corsini},
  E.~M. 2010, \aap, 521, A71

\bibitem[{{Merrifield} \& {Kuijken}(1999)}]{mer_kui_99}
{Merrifield}, M.~R., \& {Kuijken}, K. 1999, \aap, 345, L47

\bibitem[{{Mundell} \& {Shone}(1999)}]{mun_etal_99a}
{Mundell}, C.~G., \& {Shone}, D.~L. 1999, \mnras, 304, 475

\bibitem[{{Newton} \& {Emerson}(1977)}]{new_eme_77}
{Newton}, K., \& {Emerson}, D.~T. 1977, \mnras, 181, 573

\bibitem[{{Nieten} {et~al.}(2006){Nieten}, {Neininger}, {Gu{\'e}lin},
  {Ungerechts}, {Lucas}, {Berkhuijsen}, {Beck}, \&
  {Wielebinski}}]{nieten_etal_06}
{Nieten}, C., {Neininger}, N., {Gu{\'e}lin}, M., {et~al.} 2006, \aap, 453, 459

\bibitem[{{Opitsch}(2016)}]{opitsc_16}
{Opitsch}, M. 2016, PhD thesis, LMU Munich, Germany

\bibitem[{{Opitsch} {et~al.}(2018){Opitsch}, {Fabricius}, {Saglia}, {Bender},
  {Bla{\~n}a}, \& {Gerhard}}]{opitsc_etal_18}
{Opitsch}, M., {Fabricius}, M.~H., {Saglia}, R.~P., {et~al.} 2018, \aap, 611,
  A38

\bibitem[{{Pettitt} {et~al.}(2020){Pettitt}, {Dobbs}, {Baba}, {Colombo},
  {Duarte-Cabral}, {Egusa}, \& {Habe}}]{pet_etal_20}
{Pettitt}, A.~R., {Dobbs}, C.~L., {Baba}, J., {et~al.} 2020, \mnras, 498, 1159

\bibitem[{{Reynaud} \& {Downes}(1998)}]{rey_dow_98}
{Reynaud}, D., \& {Downes}, D. 1998, \aap, 337, 671

\bibitem[{{Roberts} {et~al.}(1979){Roberts}, {Huntley}, \& {van
  Albada}}]{rob_etal_79}
{Roberts}, W.~W., J., {Huntley}, J.~M., \& {van Albada}, G.~D. 1979, \apj, 233,
  67

\bibitem[{{Roberts}(1969)}]{robert_69}
{Roberts}, W.~W. 1969, \apj, 158, 123

\bibitem[{{Ruoyi} \& {Haibo}(2020)}]{ruo_hai_20}
{Ruoyi}, Z., \& {Haibo}, Y. 2020, \apjl, 905, L20

\bibitem[{{Saglia} {et~al.}(2018){Saglia}, {Opitsch}, {Fabricius}, {Bender},
  {Bla{\~n}a}, \& {Gerhard}}]{sag_etal_18}
{Saglia}, R.~P., {Opitsch}, M., {Fabricius}, M.~H., {et~al.} 2018, \aap, 618,
  A156

\bibitem[{{Sandage}(1961)}]{sandag_61}
{Sandage}, A. 1961, {The Hubble Atlas of Galaxies} ({Carnegie Inst.,
  Washington})

\bibitem[{{Sandage} \& {Bedke}(1994)}]{san_bed_94}
{Sandage}, A., \& {Bedke}, J. 1994, {The Carnegie atlas of galaxies} ({Carnegie
  Inst., Washington})

\bibitem[{{Stark}(1977)}]{stark_77}
{Stark}, A.~A. 1977, \apj, 213, 368

\bibitem[{{Stone} {et~al.}(2020){Stone}, {Tomida}, {White}, \&
  {Felker}}]{sto_etal_20}
{Stone}, J.~M., {Tomida}, K., {White}, C.~J., \& {Felker}, K.~G. 2020, \apjs,
  249, 4

\bibitem[{Virtanen {et~al.}(2020)Virtanen, Gommers, Oliphant, Haberland, Reddy,
  Cournapeau, Burovski, Peterson, Weckesser, Bright, {van der Walt}, Brett,
  Wilson, Millman, Mayorov, Nelson, Jones, Kern, Larson, Carey, Polat, Feng,
  Moore, {VanderPlas}, Laxalde, Perktold, Cimrman, Henriksen, Quintero, Harris,
  Archibald, Ribeiro, Pedregosa, {van Mulbregt}, \& {SciPy 1.0
  Contributors}}]{2020SciPy-NMeth}
Virtanen, P., Gommers, R., Oliphant, T.~E., {et~al.} 2020, Nature Methods, 17,
  261

\bibitem[{{Weiner} {et~al.}(2001{\natexlab{a}}){Weiner}, {Sellwood}, \&
  {Williams}}]{wei_etal_01b}
{Weiner}, B.~J., {Sellwood}, J.~A., \& {Williams}, T.~B. 2001{\natexlab{a}},
  \apj, 546, 931

\bibitem[{{Weiner} {et~al.}(2001{\natexlab{b}}){Weiner}, {Williams}, {van
  Gorkom}, \& {Sellwood}}]{wei_etal_01a}
{Weiner}, B.~J., {Williams}, T.~B., {van Gorkom}, J.~H., \& {Sellwood}, J.~A.
  2001{\natexlab{b}}, \apj, 546, 916

\bibitem[{{Williams} {et~al.}(2015){Williams}, {Dalcanton}, {Dolphin}, {Weisz},
  {Lewis}, {Lang}, {Bell}, {Boyer}, {Fouesneau}, {Gilbert}, {Monachesi}, \&
  {Skillman}}]{wil_etal_15}
{Williams}, B.~F., {Dalcanton}, J.~J., {Dolphin}, A.~E., {et~al.} 2015, \apj,
  806, 48

\bibitem[{{Z{\'a}nmar S{\'a}nchez} {et~al.}(2008){Z{\'a}nmar S{\'a}nchez},
  {Sellwood}, {Weiner}, \& {Williams}}]{zan_etal_08}
{Z{\'a}nmar S{\'a}nchez}, R., {Sellwood}, J.~A., {Weiner}, B.~J., \&
  {Williams}, T.~B. 2008, \apj, 674, 797

\bibitem[{{Zaritsky} \& {Lo}(1986)}]{zar_lo_86}
{Zaritsky}, D., \& {Lo}, K.~Y. 1986, \apj, 303, 66

\bibitem[{{Zurita} {et~al.}(2004){Zurita}, {Rela{\~n}o}, {Beckman}, \&
  {Knapen}}]{zur_etal_04}
{Zurita}, A., {Rela{\~n}o}, M., {Beckman}, J.~E., \& {Knapen}, J.~H. 2004,
  \aap, 413, 73

\end{thebibliography}
\appendix
\twocolumngrid

\clearpage
\section{Tests of the Edge-detection Algorithm in Identifying Velocity Jump Features}
\label{appendix:A}
Our procedure mainly follows the edge-detection algorithm in \citet{canny_86}, modified to detect step-function shaped and $\delta$-function shaped velocity jumps on PVDs. The main idea is to convolve data with the derivative of Gaussian operator 

\begin{equation}
y = -\dfrac{x}{s^2}\exp{\dfrac{-x^2}{2\sigma^2}}.
\end{equation}

\noindent Here $s = 3$ is a scaling factor and $\sigma = 4$ represents the dispersion of Gaussian.

\begin{figure}[ht]
\includegraphics[width=\columnwidth]{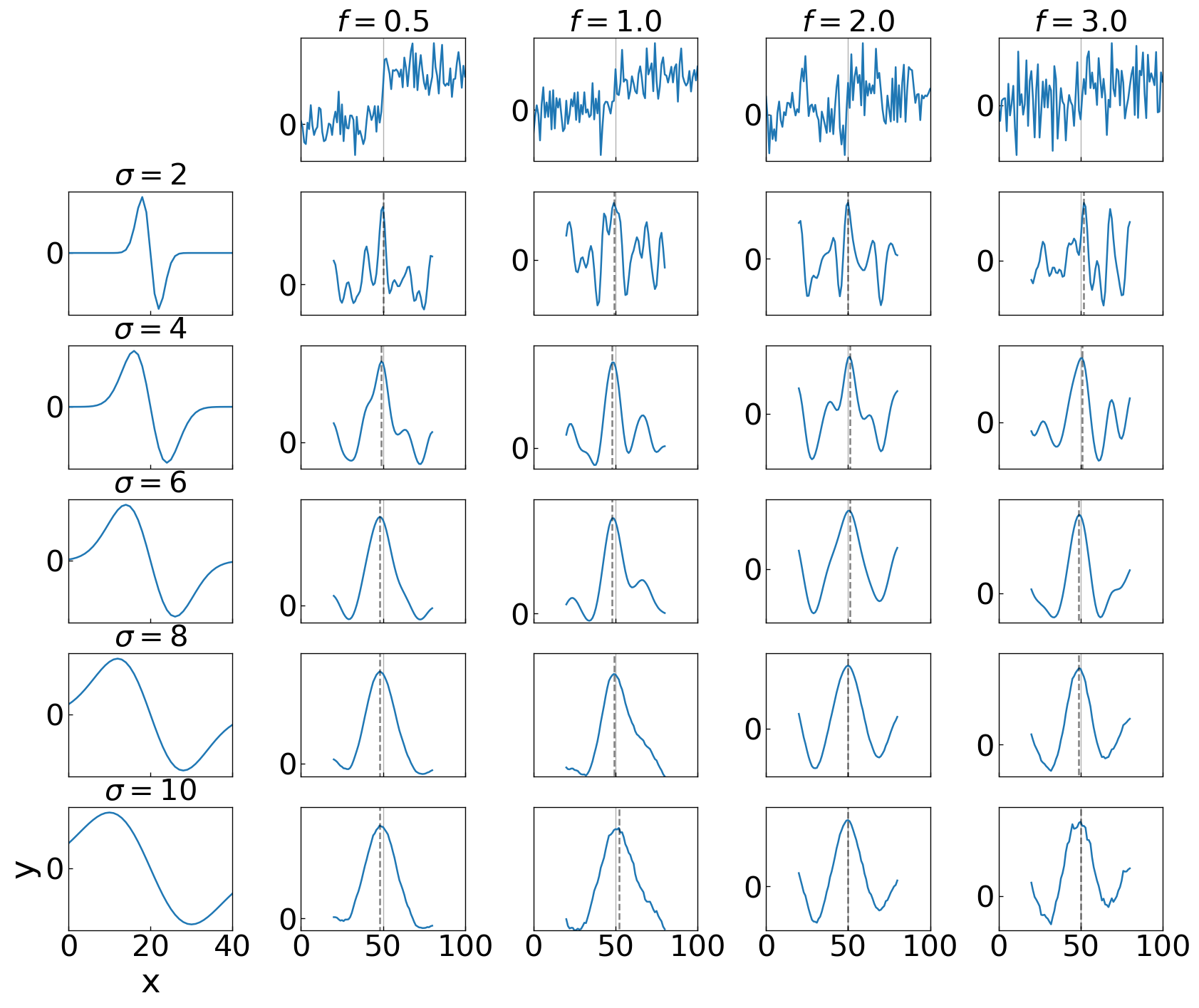}
\caption{Convolution results of step-function shaped jump features with the derivative of Gaussian operator. The top raw indicates the input mock data with jumps at $x = 50$. Gaussian noises are added on the samples with increasing amplitude from left to right columns. $f$ represents the ratio of Gaussian noise amplitude to the jump amplitude. The left column presents the derivative of Gaussian operator with different $\sigma$. Vertical solid lines represent the true jumps at $x = 50$, and vertical dashed lines indicate the identified jumps positions.}
\label{fig:step_shape_jump_identification}
\end{figure}

\begin{figure}[ht]
\includegraphics[width=\columnwidth]{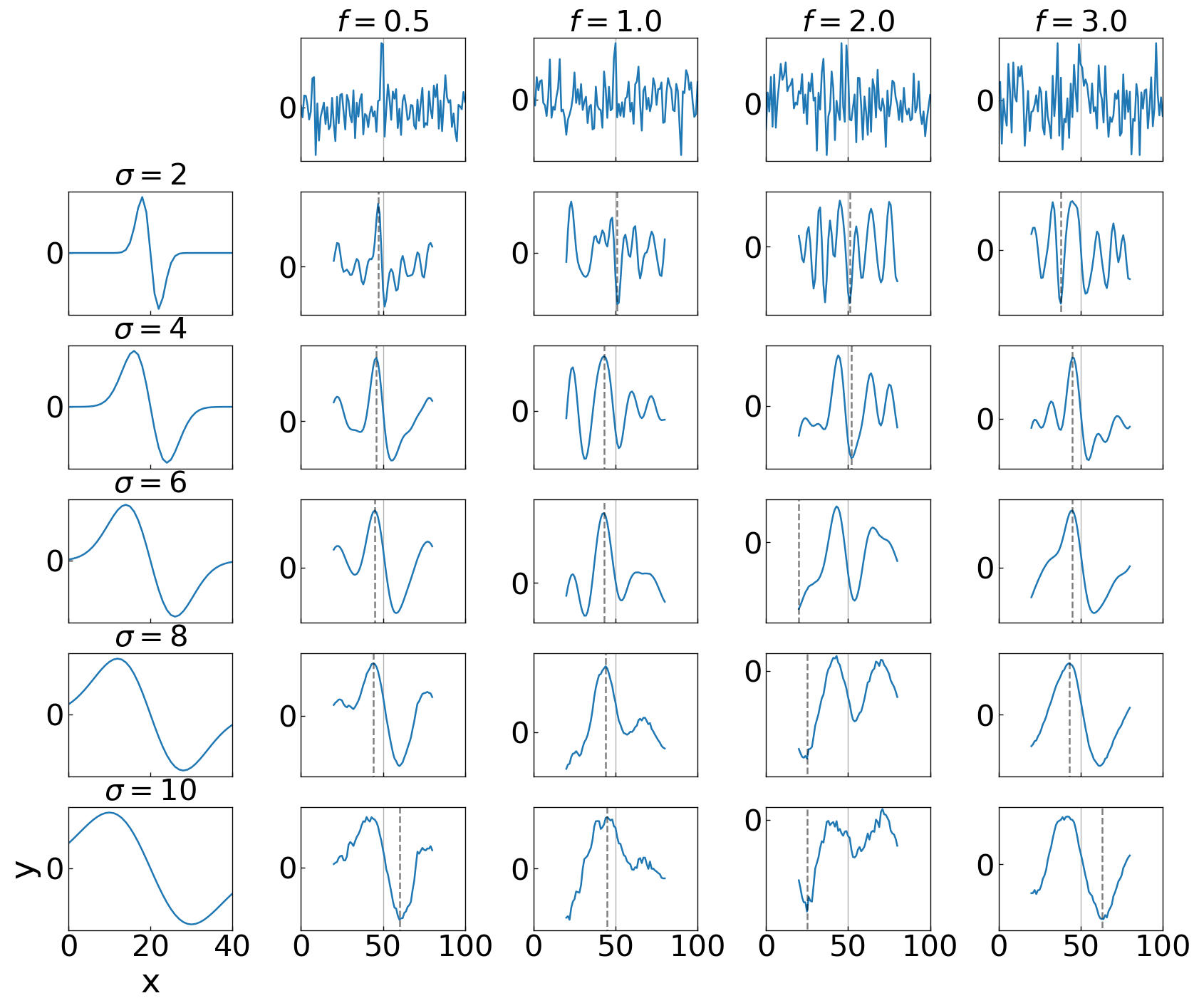}
\caption{Same as in Fig.~\ref{fig:step_shape_jump_identification} but for $\delta-$function shaped jump features.}
\label{fig:delta_shape_jump_identification}
\end{figure}

The maximum (or minimum) of the convolution result of the data and the derivative of Gaussian operator highlights the position of the velocity jump features. In Figs.~\ref{fig:step_shape_jump_identification} and \ref{fig:delta_shape_jump_identification} we test the derivative of Gaussian operator with different $\sigma$ to identify jump features in cases of four different amplitude of noises. $f$ represents the ratio of the Gaussian noise amplitude to the amplitude of velocity jumps. Here we test the cases of $f$ = 0.5, 1, 1.5 and 2. Dashed lines indicate the identified position of jump features. In general operators with a small $\sigma$ tend to highlight more local fluctuations of the sample (more peaks on the convolution result). For step-function shaped jump features, the edge-detection algorithm works well for all $\sigma$ values. A small shift of the identified position is found for larger $\sigma$ at small noises. However, for noisier data it seems larger $\sigma$ have better performance. For the $\delta-$function shaped jump features a smaller $\sigma$ is preferred for it gives more precise positions of jump features. As the noise amplitude increases, the peaks caused by noises become larger for $\sigma = 2$ and one could mistake them as the jump positions. For a balance between the stability and accuracy of edge-detection, we adopt $\sigma = 4$ in this work. 

\clearpage
\section{PVDs in slits more perpendicular to the shocks}
\label{appendix:B}
\begin{figure}[ht]
\includegraphics[width=\columnwidth]{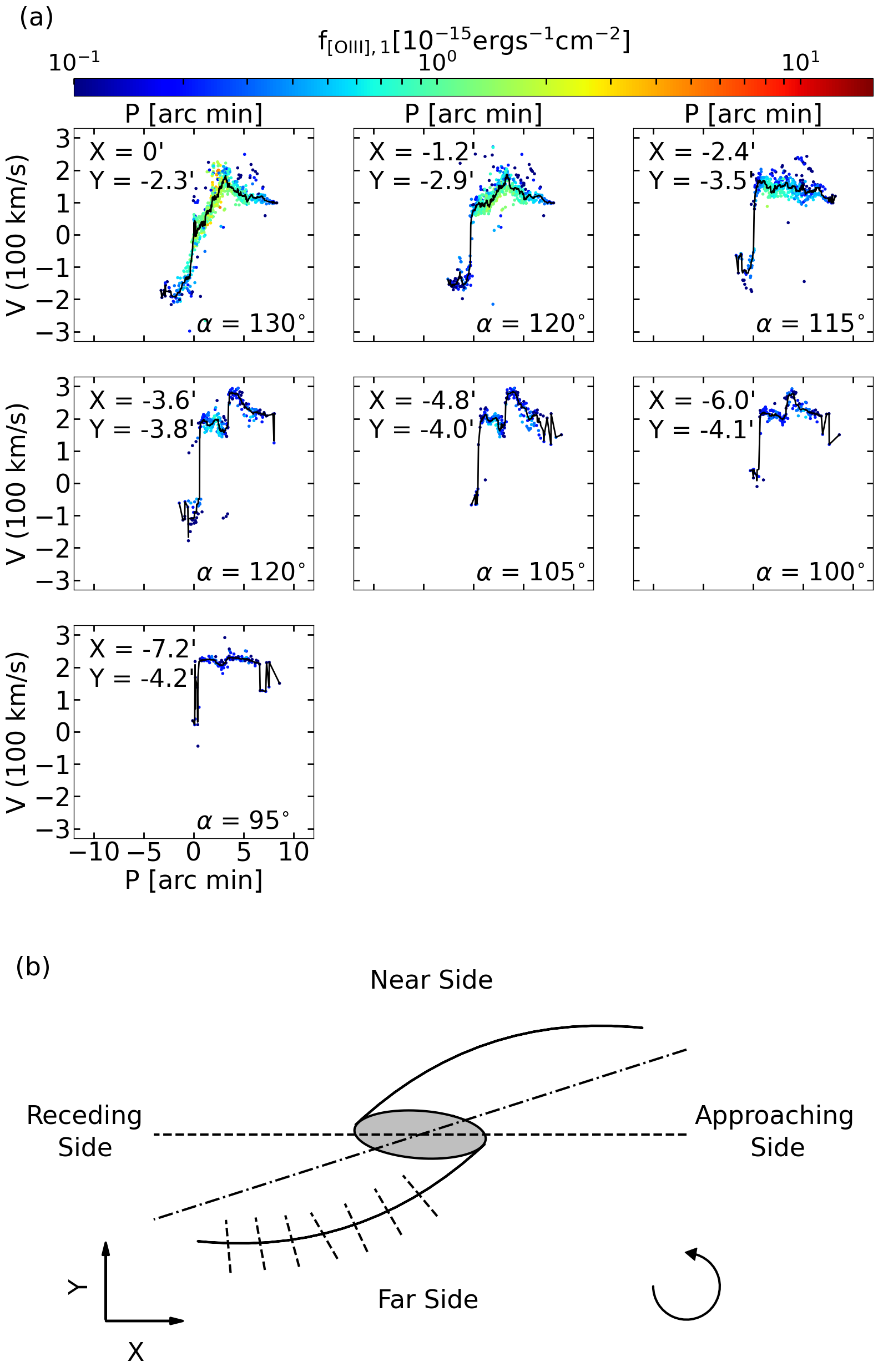}
\caption{\textit{Upper panel}: PVDs of $\firstcomponent$ color coded with flux density on the far side of M31. Slits are positioned roughly perpendicular to shock fronts to show the clearest velocity jump features. $X$ and $Y$ represent the positions where we put the centers of slits. $\alpha$ represents the angle of slits with respect to the disk major axis. Black curves indicate the boxcar smoothed results of $\firstcomponent$. \textit{Bottom panel}: Schematic plots illustrate the orientations and positions of slits we set in respect of the shocks. All of the slits have a width of $1.2\arcmin$.}
\label{fig:PVDs_OIII_far_side_appendix}
\end{figure}

\begin{figure}[!t]
\includegraphics[width=\columnwidth]{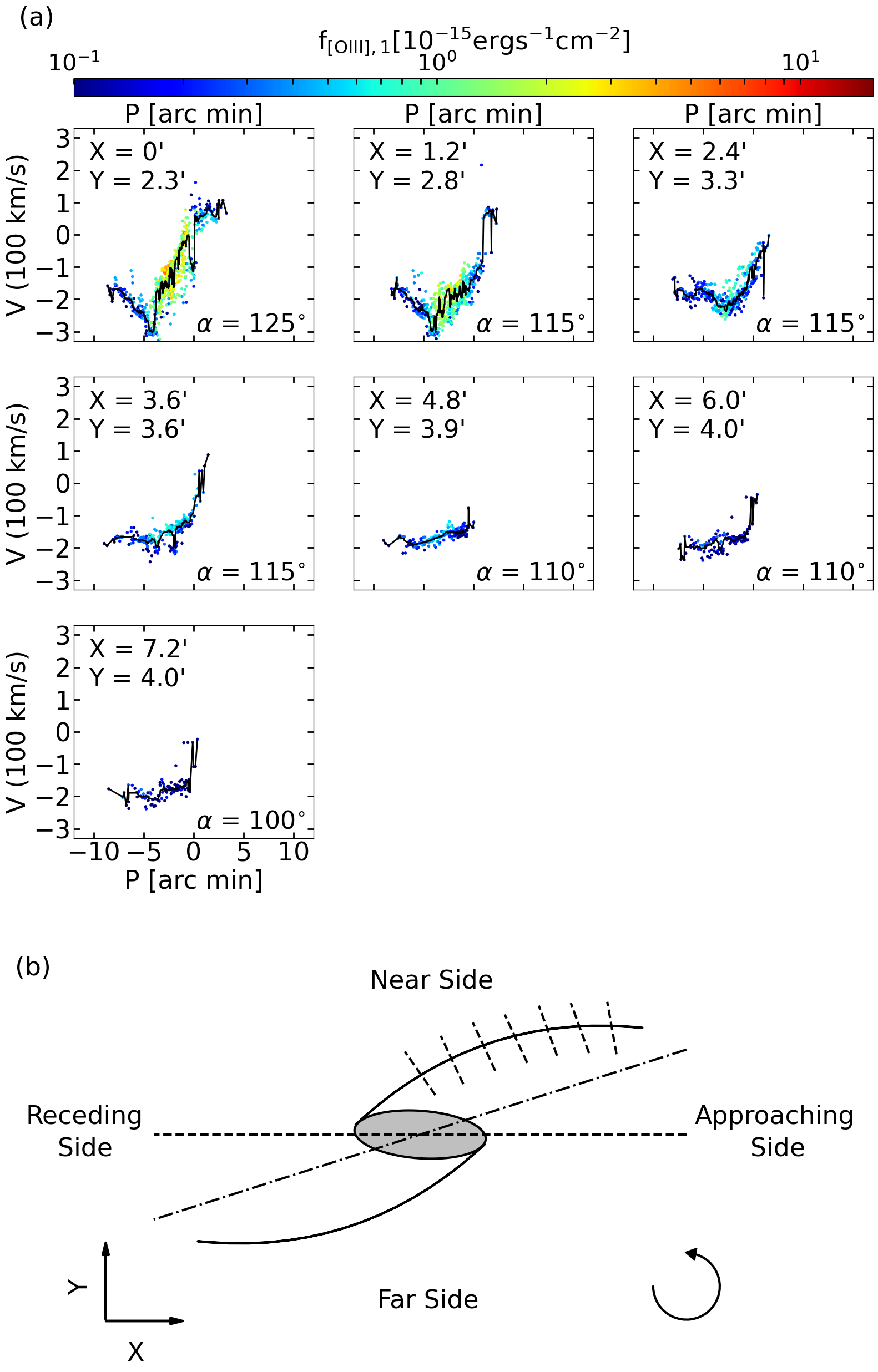}
\caption{Same as in Fig.~\ref{fig:PVDs_OIII_far_side_appendix} but on the near side.}
\label{fig:PVDs_OIII_near_side_appendix}
\end{figure}

\begin{figure}[!t]
\includegraphics[width=\columnwidth]{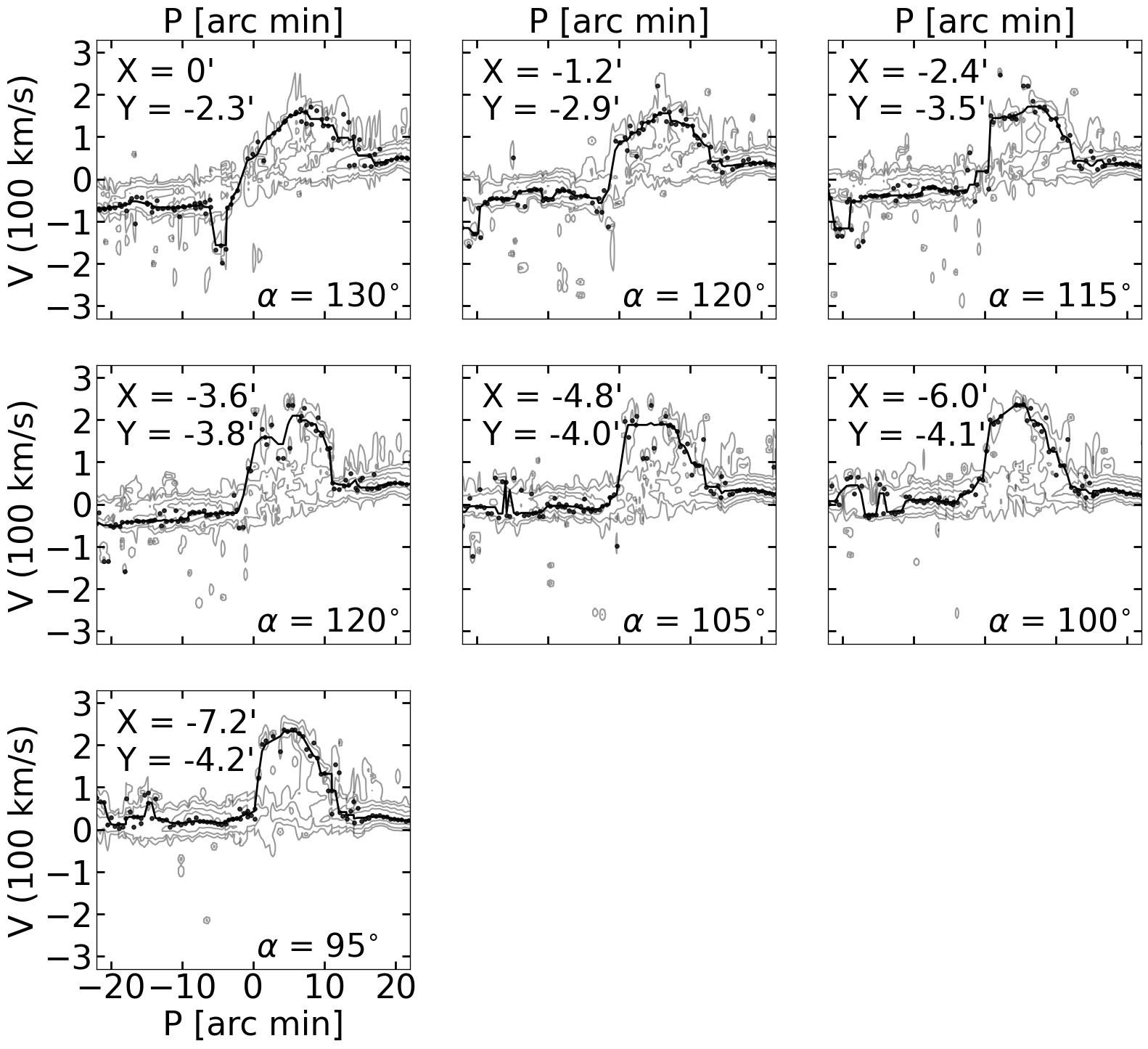}
\caption{PVDs of $\HI$ on the far side of M31. Contours indicate the integrated emission of $\HI$. Slits are positioned roughly perpendicular to shock fronts as in Fig.~\ref{fig:PVDs_OIII_far_side_appendix}. Black points indicate the velocities of the main component of $\HI$. Black curves represent the boxcar smoothed results of the black points.}
\label{fig:PVDs_HI_far_side_appendix}
\end{figure}

\begin{figure}[ht]
\includegraphics[width=\columnwidth]{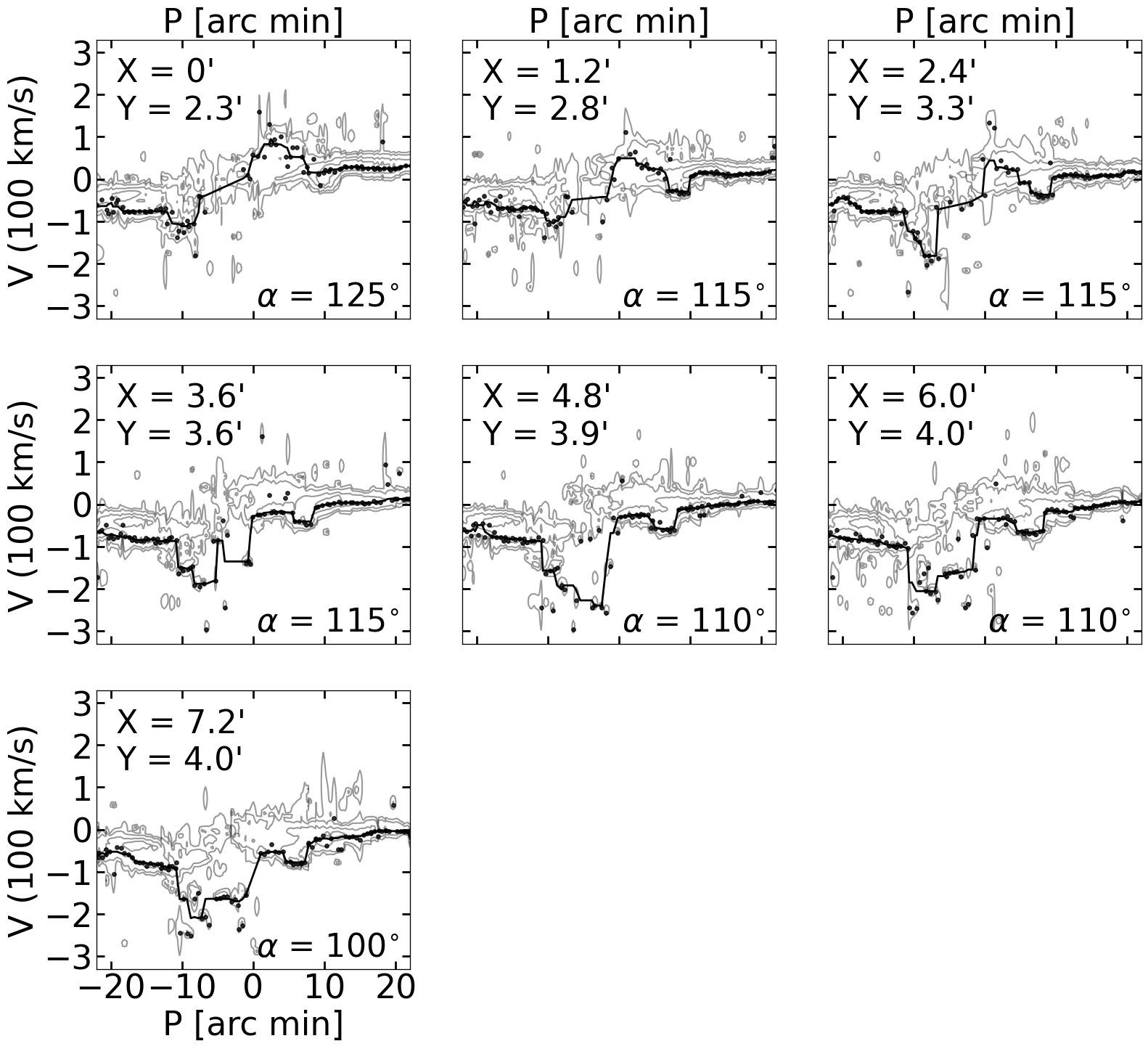}
\caption{Same as in Fig.~\ref{fig:PVDs_HI_far_side_appendix} but on the near side.}
\label{fig:PVDs_HI_near_side_appendix}
\end{figure}

Here we show the PVDs of $\firstcomponent$ and $\HI$ in slits nearly perpendicular to the shock fronts to show the clearest shock features. Figs.~\ref{fig:PVDs_OIII_far_side_appendix} and \ref{fig:PVDs_OIII_near_side_appendix} present the PVDs of $\firstcomponent$ on the far and near side of M31, respectively. Figs.~\ref{fig:PVDs_HI_far_side_appendix} and \ref{fig:PVDs_HI_near_side_appendix} present the PVDs of $\HI$ on the far and near side of M31, respectively. Overall amplitude and the sharpness of the shock features are similar to those in \S\ref{sec:result}. 

\clearpage

\end{document}